\documentclass[print]{aa} 

\usepackage{graphicx}   
\usepackage{amsmath}    

\usepackage{placeins}

\usepackage{txfonts}
\usepackage{multirow}
\usepackage{colortbl}
\usepackage{hyperref}  
\hypersetup{colorlinks=true,linkcolor=[rgb]{1.,0.2,0.2},citecolor=[rgb]{0.1,0.4,1.},filecolor=[rgb]{0.7,0.2,0.2},urlcolor=[rgb]{0.7,0.2,0.2}}

\usepackage{color}
\definecolor{blue}{rgb}{0., 0., 1}
\definecolor{lightblue}{rgb}{0.1,0.4,1.}

\usepackage{soul}

\def\teff{$T\rm_{eff}$}

\newcommand{\kms}{$\rm km\, s ^{-1}$}
\newcommand{\mygi}{MyGIsFOS}

\newcommand{\msun}{\mathrm{M}_\odot}
\DeclareRobustCommand{\ion}[2]{\textup{#1\,\textsc{\lowercase{#2}}}}

\usepackage[dvipsnames]{xcolor}

\usepackage{soul}

\begin{document} 

\title{The accretion history of the Milky Way\\ IV. Hints of recent star formation in Milky Way dwarf spheroidal galaxies} 
\titlerunning{Recent star formation in MW dSphs}
\authorrunning{Y.\ Yang et al.} 


\author{Yanbin Yang\inst{\ref{ins1}} \fnmsep\thanks{E-mail: \href{mailto:yanbin.yang@obspm.fr}{yanbin.yang@obspm.fr}}
\and Elisabetta Caffau\inst{\ref{ins1}}
\and Piercarlo Bonifacio\inst{\ref{ins1}}
\and Fran{\c c}ois Hammer\inst{\ref{ins1}} 
\and Jianling Wang\inst{\ref{ins2}} 
\and Gary A. Mamon\inst{\ref{ins3}}
}
\institute{ GEPI, Observatoire de Paris, Universit\'e PSL, CNRS, Place Jules Janssen, 92190 Meudon, France \label{ins1}
\and CAS Key Laboratory of Optical Astronomy, National Astronomical Observatories, Beijing 100101, China  \label{ins2}
\and Institut d’Astrophysique de Paris (UMR7095: CNRS \& Sorbonne Université), 98 bis Bd Arago, 75014, Paris, France \label{ins3}
}

\date{Received 24 June 2024; accepted 14 September 2024 }

 
\abstract{Dwarf spheroidal galaxies are known to be dominated by old stellar populations. This has led to the assumption that their gas-rich progenitors lost their gas during their infall in the Milky Way (MW) halo at distant look-back times. 
Here, we report a discovery of a tiny but robustly detected population of possibly young ($\sim$ 1 Gyr old) and intermediate-mass 
($\rm 1.8 M_{\odot} \le M < 3 M_{\odot}$) stars in MW dwarf spheroidal galaxies. This was established on the basis of their positions in color-magnitude diagrams, after filtering out the bulk of the foreground MW  using {\it Gaia} DR3 proper motions. 
We have considered the possibility that this population is made of evolved blue stragglers. For Sculptor, it seems unlikely, because 95.5\% of its stars are older than 8 Gyr, leading to masses smaller than 0.9 M$_{\odot}$. This would only allow blue straggler masses of less than 1.8 M$_{\odot}$, which is much lower than what we observed. Alternatively, it would require the merger of three turnoff stars, which appears even more unlikely. On the other hand, the recent {\it Gaia} proper motion measurements of MW dwarf galaxies infer their low binding energies and large angular momenta, pointing to a more recent, $\le$ 3 Gyr, infall. 
Although the nature of the newly discovered stars still needs further confirmation, we find that  they are consistent with the recent infall of the
dwarf galaxies into the MW halo, when star formation occurred from the ram pressurization of their gas content before its removal by the hot Galactic corona. The abundance of this plausibly young population of stars is similar to the expectations drawn from hydrodynamical simulations. These results point to a novel origin for MW dwarf spheroidal galaxies.

}

\keywords{galaxies: abundances -- galaxies: dwarf -- galaxies: star formation}

\maketitle

\section{Introduction}

It is widely accepted that dwarf galaxies have been accreted by the Milky Way (MW) during a process of gas loss due to ram pressure stripping exerted by the Galactic hot corona. This also provides a reasonable explanation of the morphology density relation \citep{Putman2021}, whereby dwarfs inside (outside) 250 kpc from their MW or M31 host are mostly gas-free and dispersion supported (gas-rich and rotation-supported) systems, respectively. \\

However, the infall time of most MW dwarfs (those within $\sim$ 250 kpc) is still a matter of debate. For a long time, it was  accepted that the cutoff of dwarf star formation history (SFH) provides a good clock to estimate their infall time. This argument is mostly limited to classical dwarf spheroidal galaxies (dSphs) for which the stellar population (especially of the red-giant branch, RGB) is sufficiently abundant to retrieve their SFHs. It leads to contrasting results; for instance, Sculptor \citep{DeBoer12}, Ursa Minor, Sextans \citep{Bettinelli2018, Lee2009,Carrera2002}, and Draco \citep{Mapelli2007} showed a sharp decrease of their SFHs 8 to 10 Gyr ago. Meanwhile, Fornax \citep{Saviane2000,DeBoer12Fornax}, Carina \citep{DeBoer14,Monelli2003,Weisz2014}, and Leo I \citep{Ruiz-Lara2021} showed extended SFHs including toward very recent epochs (0.1 to 2 Gyr ago).  Intermediate-to-young stars have been detected in the central region of LeoII \citep{Komiyama2007}. Canes Venatici~I, having a stellar mass similar to that of Carina and of UMi,
also shows star formation activity about 1.5--2 Gyr ago \citep{Weisz2014,Martin2008}. \\

Concerning more in-depth studies of stellar population, the notion that dSphs are mainly made up of old stars has been met with a broad consensus, based on a clearly defined turnoff (TO) of an old population, visible in the color-magnitude diagrams (CMDs) of all dSphs \citep[see][and references therein]{tolstoy09}. The possible presence of a young
population in dSph galaxies, on the other hand, is highly controversial. For example, while there is a consensus on the
dominance of an old population in the Draco dSph, \citet{Aparicio2001} claimed the detection of a young population of 2--3 Gyr. On the other hand, \citet{Mapelli2007} provided circumstantial evidence (mainly based
on the spatial distribution of ``blue plume,'' BP, stars) of an absence of young populations; in addition, these authors posited that BPs are indeed blue straggler stars (hereafter, BSSs), although they clearly mention that the presence of a young population suggested by \citet{Aparicio2001} cannot be ruled out. 
The claims of a young population in dSph galaxies hinge on the ubiquitous ``blue plumes'' (BP) observed in the CMDs of all dSph galaxies \citep[see][]{Momany2007, santana13}. 
These BP stars appear as an extension to the blue of the main sequence (MS), beyond the TO of the old population.
The claims of the existence of young populations
have also been based on another feature often seen in
CMDs: a vertical sequence of stars bluer than
the RGB, which extends (in certain cases) to magnitudes
brighter than the horizontal branch (HB). This sequence is usually
called the ``vertical clump'' \citep[VC,][]{Gallart2005}
or ``yellow plume'' \citep[e.g.,][]{Gull2008} or even
``blue loop'' \citep{DeBoer12}. 
The location of this vertical sequence of blue stars in the CMD is consistent with stars of mass in the range $ 1.5 \le M/\msun \le 4 $
and ages younger than 4 Gyr, in the phase of core-helium burning (CHeB).
We refer to these stars as CHeBs,
although some sub-giants stars in the same mass range, can occupy the same region
in the CMD.
The interpretation of these features in CMDs is ambiguous: while the BP and the CHeB can be interpreted
as MS and CHeB phases of a young population,
they can also be interpreted as BSSs and  evolved
BSSs (in the CHeB phase).\\

The above discussion leads to some ambiguity on the SFH of dSphs, and, thus, their related infall times. However, the accurate proper motions of the MW dSphs from the {\it Gaia} EDR3 \citep{Li2021,Battaglia2022} have revolutionized our understanding of the dSph orbital motions, providing accurate values of orbital energies for a given MW mass model. In the accepted hierarchical scenario of structure formation, the most recent newcomers typically have smaller binding energies than  satellites that entered  at early epochs \citep{Gott1975}. Such a behavior has been confirmed with cosmological simulations \citep{Rocha2012,Santistevan2023},  demonstrating a tight correlation between the infall lookback time and the binding energy. \citet{DSouza2022} retrieved the same correlation for most of the simulated halos of the ELVIS suite \citep{Garrison-Kimmel2014}, finding it was corrupted in only 3 of 48 halos having a very active merger history, which unlikely applies to the relatively quiet MW \citep{Hammer2007}. In particular, \citet[hereafter Paper I, see their Fig. 6]{hammer23} confirmed this relationship for the MW, using the globular clusters associated to the bulge, as well as those associated with the Kraken, {\it Gaia}-Sausage-Enceladus (GSE), and Sgr, which   merged with the MW early on. The latter events are associated to much larger binding energies than that of dSphs.  In other words, events that took place 8--10 (GSE) and 4--6 (Sgr) Gyr ago, are six and three times more bound than dSphs, respectively. 
Paper~I presented the conclusion that MW dSphs are likely newcomers, arriving into the MW halo during the past 3 Gyr. The main advantage of the above approach is that it does not depend on the assumed total mass value of the MW, since it is only based on the energy difference with well-known events (Kraken, GSE, and Sgr). Other studies based on {\it Gaia} proper motions \citep{Fritz2019,Battaglia2022} may offer different conclusions, depending on their assumed choice for the MW mass (see a discussion in \citealt{Li2021}), which is also an important matter of debate \citep{Jiao2023,Ou2024}. \\

A timescale of $\le$ 3 Gyr is too short for MW dSphs to make more than one orbit because they lie at large distances. For most of them, they would be at their first entry in the MW halo, 
similar to the Magellanic Clouds \citep[MCs,][]{Kallivayalil2013}. A recent infall history for dSphs means that they may have been gas-rich less than 3 Gyr ago, as are the dwarf irregulars that dominate the dwarf population of the Local Group at distances greater than 250 kpc from the MW or M31 \citep[see e.g.,][]{Grcevich2009,Putman2021}. Ram pressure stripping of their gas by the hot gas corona of the MW is understood to have caused their transformation into dSphs \citep[e.g.,][]{Mayer2006,Yang2014}, to which we should add the impact of tidal shocks (\citealt{Aguilar1988}; \citealt{Hammer2024}, hereafter Paper~II). This mechanism has been explored by \citeauthor{Wang2024} (\citeyear{Wang2024}, hereafter Paper~III), who used hydrodynamical simulations to successfully reproduce the morphologies, internal kinematics, and orbital properties of, for instance, Sculptor. If dSphs fell into MW within the last 3~Gyr as proposed in Paper~I, the transformation from dIrrs to dSphs would have occurred rapidly and violently (see Paper~III). In particular, gas is expected to be compressed before its removal,
leading to the formation of young stars \citep{Bothun&Dressler86}, whose (small) fraction is predicted in Paper~III. \\

The goal of this paper is to verify whether or not there is a population of young, less than 3 Gyr-old stars, in the dSphs but Sgr. The main text focuses on the Sculptor dSph properties, while most properties of other dSphs are presented in the Appendices.
In Sect.~2, we describe our strategy to identify young stellar population in the dSphs. 
In Sect.~3, we describe the analysis and results based on the {\it Gaia} photometry.
Section~4 presents the analysis of a spectroscopic sample of Sculptor stars. 
 Our discussion and conclusion are given in Sect.~5 and 6, respectively.

\begin{figure}
\centering
\resizebox{8.5cm}{!}{\includegraphics{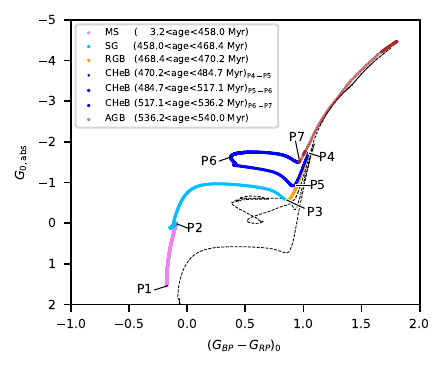}}
\caption{Example of the PARSEC evolutionary track of an intermediate-mass star ($m=2.4 M_{\odot}$ and 
${\rm [Fe/H]}=-1.30$) in the photometric system of {\it Gaia}. Numbers P1--7 correspond to the starting point of each evolutionary phase encoded by colors, as explained in the legend (together with the age interval of each phase indicated). The subscript ``0"  in the axis labels denotes that no reddening has been applied.
The dashed line represents an evolutionary track (within 665 Myr) of a BSS of 1.57~$M_\odot$ and [Fe/H]$=-2.3$, which is formed by the merger of two MS stars of 0.8~$M_\odot$ \citep{Sills2009}.
}
\label{phases}
\end{figure}

\section{Methodology}

The MS stars that are most proliferous in number would be the ideal stellar population for probing recent star formation. 
However, to study the MS star in dSphs is very difficult because of their mix with BSSs in the BP region of the CMD. 
As an alternative probe of recent star formation, we propose investigating
intermediate-mass stars (between 1.5 to 8 $M_{\odot}$) when they are in the sub-giant (SG) phase and the CHeB phase. In Figure~\ref{phases}, we show a PARSEC  stellar evolutionary track
\citep{bressan12}  of an intermediate-mass star. 
We may notice that after leaving the main sequence (P1--P2)
, the star rapidly passes through the SG phase (P2--P3) on a Kelvin-Helmholtz or thermal time scale (10 Myr here), and then enters the RG phase at P3. 
At the tip of RGB (P4), it ignites helium burning in the core, when phases P4 to P7 cause the star to perform a ``blue loop'' in the CMD,
before reaching the asymptotic giant branch (AGB) phase (P7). 
All phases before P7 take 536 Myr, which include almost 90\% of the time in the MS and about 10\% in the SG and the CHeB phases. Depending on the mass, age, and metallicity of stars, the shape of the blue loop changes considerably  over a wide region of CMD.
When considering all evolutionary tracks of intermediate-mass stars, the CHeB phases overlap 
with part of the SG phases in CMD and occupy the region above the HB of the old
stellar population and bluer than the RGB stars. 
The CHeB phase may last for a few tens of Myr, depending on the masses and metallicities of stars. 

CHeB stars are young, of typically less  than 2~Gyr. 
For example, the Classical Cepheids are the stars in CHeB phase when they are crossing the instability strip \citep{Christy1966,Fernie1990,Maran1991book}.  In most of the dSphs, a handful sample of Anomalous Cepheid (AC) has been noticed \citep[][and references therein]{Mateo1998,Kinemuchi2008,Monelli2022,Soszynski2020}. On the other hand, it is unclear whether they are young stars or rejuvenated stars that were formed by binaries (i.e., evolved BSSs). 

To identify the full CHeB population in dSphs is challenging because there are not many of them and they can be highly contaminated by foreground MW stars due their locations in the CMD \citep[see e.g.,][]{DeBoer2011}. 
Here, we propose a different approach to select CHeB stars in MW dSphs by using the {\it Gaia} data. Since the dSphs are moving differently from most of the MW stars, we may use proper motion (PM) to distinguish their stars from those belonging to the MW. We may call this technique "PM-filter," as adopted by \citet{Yang2022}. These authors were able to suppress the MW contamination down to a level of 0.1\% in the Fornax dSph field of view. This allows to reach a background of contamination 100 times smaller than ground-based observations, which is very useful for deriving the density profile of dSphs, including Fornax. 
The use of PM for membership selection in dwarf galaxies and clusters 
is not new \citep[see e.g.,][]{2001AJ....122.3106E}, but the quantity and quality of PMs provided by {\it Gaia}, with respect
to what was available from ground-based observations, makes it much more powerful and precise.
CHeB stars in some of the MW dSphs are brighter than 20 mag in the {\it Gaia} $G$-band. Thus, the removal of the foreground MW stars is expected to be efficient thanks to the high quality data of {\it Gaia}.

Furthermore, the spectroscopic confirmation of CHeB candidates is helpful to confirm their membership, as well as to provide metallicity in order to resolve the degeneracy between mass, age, and metallicity. Here, we have chosen Sculptor as the best target to perform spectroscopic analysis for CHeB candidates, because spectra for these stars
are available in the ESO archive.
Sculptor is often considered as an archetype for an old dSph, without significant star formation in the past 5-7~Gyr \citep{DeBoer12}.

\section{Search for CHeB populations in dSphs}
\label{sec:chebdetection}

\subsection{{\it Gaia} EDR3 and PM-selected member sample of dSphs}
\label{sec:Gaiadata}
In this subsection we focus in identifying dSph stars using a selection from their PMs using {\it Gaia} EDR3 data. 
Our study focuses on the classical dSphs listed in Table~\ref{tab:par}. We have excluded Leo~I, and Leo~II because they are too far to be reached by {\it Gaia} for our analysis. Sgr has been also removed since it has a very different infall history compared to other classical dSphs (see, e.g., Paper I). 
For each dSph we have downloaded all {\it Gaia} data within a 10-degree radius from the main table of the {\it Gaia} EDR3 archive\footnote{{\it Gaia} Archive at \url{https://gea.esac.esa.int/archive/}}, with the combination of the following conditions: 
1) not \texttt{duplicated\_source}; 
2) not QSO;
3) with color \texttt{bp\_rp} measured;
4) with astrometric solutions (either five parameters or six parameters);  5)
$G<20.8$;
 6) \texttt{ruwe < 1.4};
and 7) \texttt{$C^{*} < 1.0$}. Here, $C^{*}$ is the corrected \texttt{phot\_bp\_rp\_excess\_factor} introduced by \citet{Riello2021}, which is efficient for removing background galaxies using the condition we adopted here (see their Figure~$\!$21). 
To exclude QSOs, we used the sample provided by the {\it Gaia} EDR3; namely, the table \texttt{gaiaedr3.agn\_cross\_id}.

Following the method introduced by \citet{Yang2022}, as well as their coordinate definition centered on object, we applied the PM-filter to the {\it Gaia} data and obtained a ``PM-selected" sample for each dSph. Below, we provide a detailed description for the Sculptor dSph. 
We have also derived their morphological parameters, such as the position angle (PA) and ellipticity, $e$, which are listed in Table~\ref{tab:par}.\\ 

\begin{figure*}
\centering
\sidecaption
\resizebox{12cm}{!}{\includegraphics{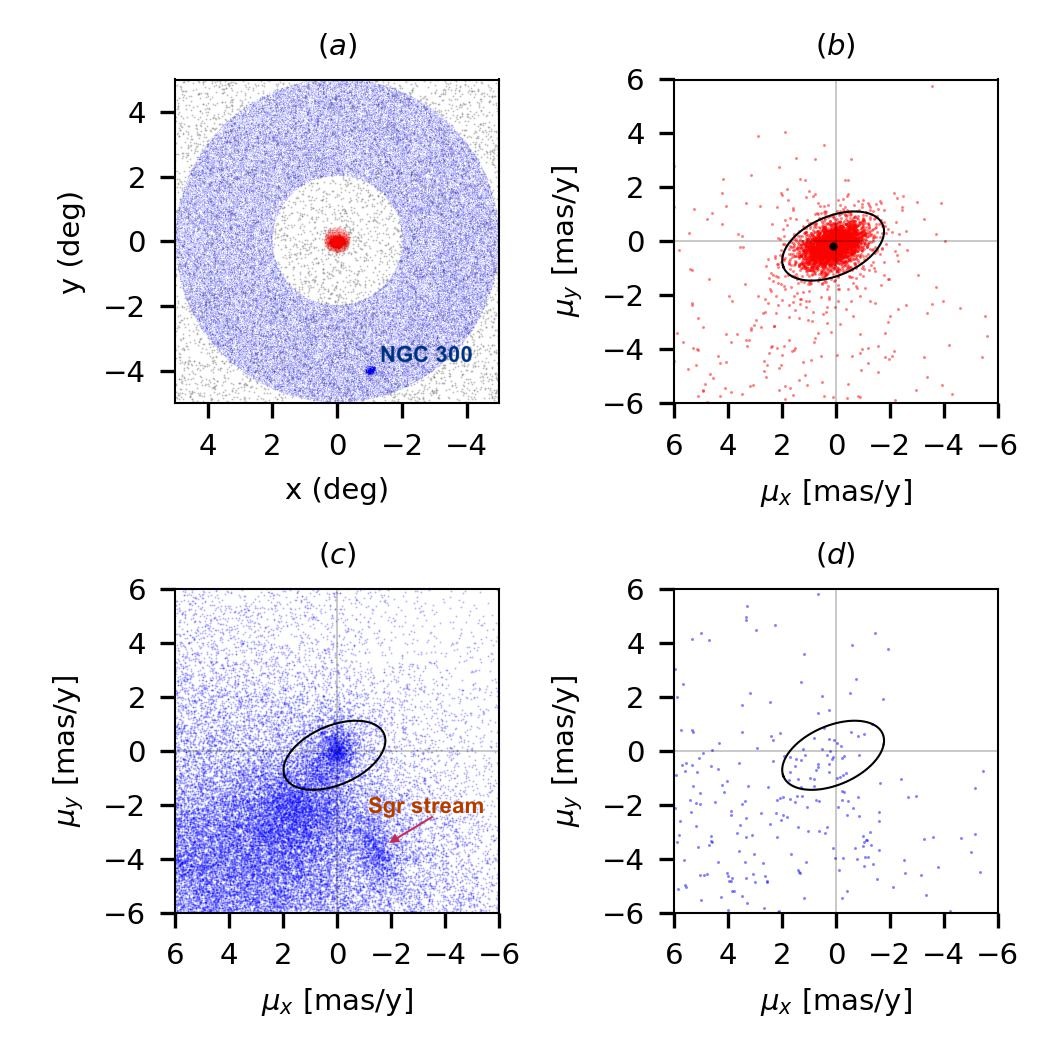}}
\caption{Diagnosis of the PM distribution of stars in the field centered on Sculptor.
Panel ($a$) shows the spatial distribution of all {\it Gaia} sources (gray dots) in a $5\times5$~sq~degree field of view, centered on Sculptor. Among these {\it Gaia} sources, we defined an object-sample (red dots) by selecting all sources within a circle of 0.4-degree radius and a control-sample (blue dots) by selecting all sources within an annulus from 2 to 5 degrees.
Panel ($b$) gives the proper motion distribution of the object-sample. Panel ($c$) provides the proper motion distribution of the control-sample. 
In panels ($b$), ($c$), and ($d$), the black eclipse (with a major-axis radius of 2.0 mas/yr) indicates the maximal region of our proper motion selection of Sculptor member candidates.
To illustrate the MW contamination inside the black ellipse, panel ($d$) shows the proper motion distribution of a sub-control-sample within an annulus from 4.0 to 4.02 degrees, which covers the same sky area as the black ellipse. }
\label{pminspect}
\end{figure*}

Here, we briefly describe the algorithms (see also Sect.~4.1 in \citealt{Yang2022}) we used to obtain the proper-motion (PM)-selected sample for Sculptor star members. The same methodology has been applied for other dSph star members.
First, we examine the proper motion distribution of the raw {\it Gaia} data, by defining an object-sample and a control-sample, as shown in Figure~\ref{pminspect} ($a$). 

The proper motion distributions of both the object-sample and the control-sample are shown in Figure~\ref{pminspect} ($b$) and ($c$), respectively.
The object-sample primarily consists of the member stars of Sculptor. 
Thus, the concentration of proper motion in Figure~\ref{pminspect} ($b$) is dominated by the member stars of Sculptor. 
The dispersion in a elliptical shape of the concentration reflects the uncertainty of proper motions, and the correlation between $\mu_x$ and $\mu_y$, respectively. 
By fitting the PM distribution of the object-sample, we obtained a black ellipse (as plotted in the panel) that encloses most of the member stars of Sculptor. It defines our selection region of member candidates in PM space.

We could improve the selection by defining a member candidate when its proper motion is consistent with the mean proper motion of Sculptor, within a tolerance of three times its own uncertainty. The later condition takes into account the fact that the uncertainties in the proper motions change with the magnitude of the sources. Therefore, brighter MW sources with better accuracy could be easily eliminated, as a result of the proper motion selection (i.e., the PM-filter).

Our primary goal is to search for unknown stellar population in dSphs. Thus, our method is relatively simple without assuming priors in colours of stars and their spatial distribution as adopted in the likelihood algorithm by \citet{Pace2019}, see also \citet{McConnachie2020} and \citet{Battaglia2022}. Our choice avoids any possible bias 
at the price of accepting a certain level of non-member contamination, which can, however, be evaluated in the subsequent analysis.

Figure~\ref{pminspect} ($d$) illustrates how efficiently the PM-filter eliminates contamination. It displays a subset of the control-sample in an annulus ranging from radii of 4.0 to 4.02 degrees, which covers the same sky area as the object-sample does. Therefore, Figure~\ref{pminspect} ($d$) illustrates that the level of contamination in the object-sample is small (0.7\%). 

\subsection{Preparation of PARSEC evolutionary tracks and BaSTI }
\label{preparationtracks}
We used PARSEC stellar evolutionary tracks\footnote{\url{https://people.sissa.it/~sbressan/CAF09_V1.2S_M36_LT/}} and isochrones\footnote{\texttt{CMD~3.7} at \url{http://stev.oapd.inaf.it/cgi-bin/cmd_3.7}} as the reference to identify CHeB star candidates, as well as to study their stellar parameters.
Using the YBC database of stellar bolometric corrections \citep{2019A&A...632A.105C}, we 
have computed magnitudes and colors in
the {\it Gaia} photometric system; namely, $G$, $G_{BP}$, and $G_{RP}$ including Galactic extinction for 
all PARSEC tracks. We prefer to ``redden'' the theoretical tracks, rather than ``de-redden'' the observed photometry,
because the extinction coefficients depend on effective temperature and gravity of the tracks. While these are perfectly well-known
for the theoretical tracks, they are not so for the observed stars. 
For each dSph, we computed the magnitudes and colors, according to its Galactic extinction (Table~\ref{tab:par}) for a full set of  PARSEC evolutionary tracks. In the following analysis, when we mention evolutionary tracks, we implicitly refer to the track library dedicated to the corresponding dSph.

In the discussion, we also used \texttt{BaSTI}\footnote{\url{http://basti-iac.oa-teramo.inaf.it/syncmd.html}} \citep{Pietrinferni2024,Pietrinferni2021,Hidalgo2018} to simulate synthetic CMD in order discuss the stellar population of dSphs. 
Although BaSTI is based on a stellar evolutionary library different from PARSEC, for our discussion of stellar population this has a marginal impact because synthetic CMD is more related to the star formation history of dSphs and the initial mass function (IMF) of star formation.

\begin{table*}
\caption{Parameters of dSphs}
\label{tab:par}
\begin{tabular}{lcccccc}
\hline\hline
      & (1) & (2) & (3) & (4) & (5) & (6)  \\
  Name & $(m-M_0)$   & $E(B-V)$   &  [Fe/H] & PA & $e$ & $r_{\rm half}$\\
      & & &  &(deg) & & (deg; kpc) \\
\hline
  Sculptor & $19.67 \pm 0.13$ & $ 0.0135\pm 0.0013$ & $(-2.5,\ -0.8)$ &95.4 & 0.28 & 0.205; 0.308\\
   Sextans & $19.89 \pm 0.07$ & $ 0.0449\pm 0.0046$ & $(-3.0,\ -0.7)$ &51.1 & 0.55 & 0.290; 0.382\\
       UMi & $19.4 \pm 0.11 $& $ 0.0283\pm 0.0047$  & $(-3.5,\ -0.7)$ &88.0 & 0.32 & 0.166;  0.240 \\
     Draco & $19.57\pm0.16$ & $ 0.0290\pm 0.0028$   & $(-3.0,\ -0.8)$ &30.5 & 0.12 & 0.295; 0.433\\
    Carina & $20.13 \pm 0.10$& $ 0.0600\pm 0.0075$   & $(-3.5,\ -0.4)$ &61.1 &  0.31 & 0.191; 0.353\\
    Fornax & $20.72 \pm 0.05$& $ 0.0251\pm 0.0041$  & $(-2.8,\ \ \ \,0.0)$ &47.3 & 0.32 & 0.307; 0.745\\
\hline
\end{tabular} \\
Column-(1): Distance modulus from \citet{Li2021} and references therein. \\
Column-(2): Galactic extinction, that are calculated by averaging the extinction map within 1 degree around each object, based on the map by \citet{Schlegel1998} via https://irsa.ipac.caltech.edu/applications/DUST/. \\
Column-(3): Metallicity range, which are taken from \citet{Kirby2011}, except for Carina (from \citet{Koch2006}).\\
Column-(4-6): Morphological parameters: Position Angle (PA), ellipticity ($e$), half-light radius, respectively. We adopted $e=1-b/a$, where $a$ and $b$ are the major and minor radius, respectively. PA and $e$ have been calculated from our PM-selected sample and $r_{\rm half}$ values have been taken from \citet{Munoz2018}.
\end{table*}


\begin{figure*}
\centering
\resizebox{18cm}{!}{\includegraphics{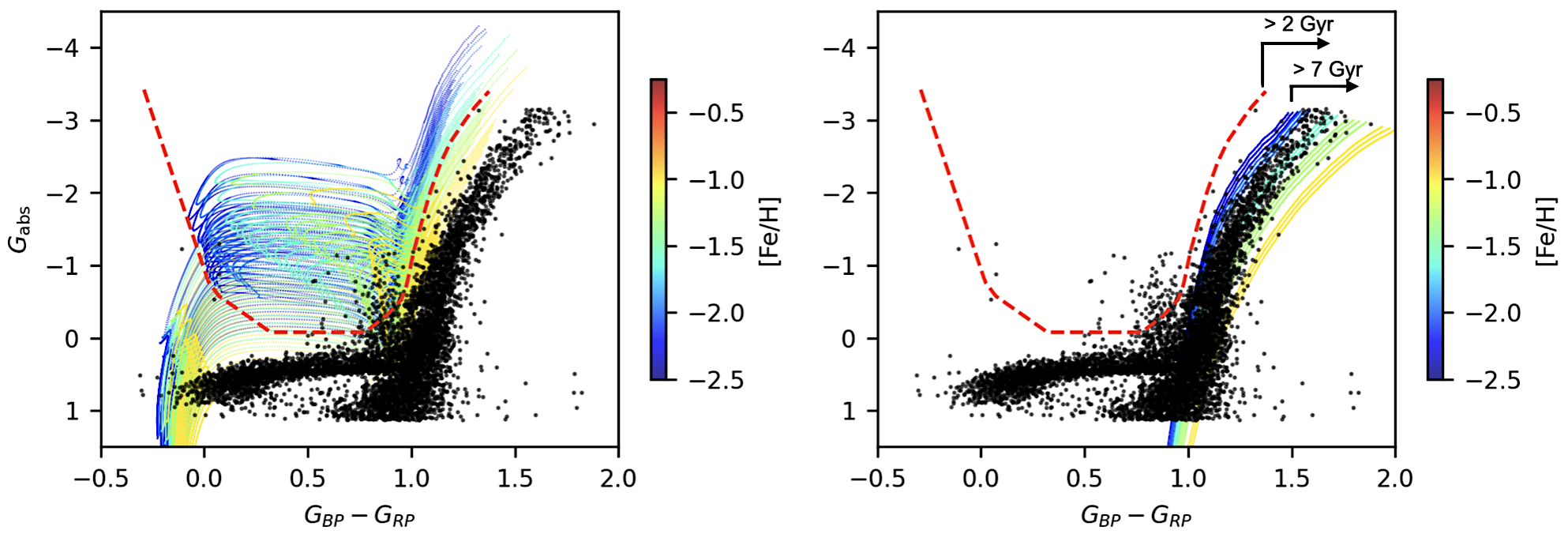}}
\caption{Sculptor stars (black points) in the CMD based absolute $G_{\rm abs}$ magnitudes and $G_{BP}-G_{RP}$ colors. The sample is selected from a region of 1.1-degree elliptical radius. 
In both panels, PARSEC stellar evolutionary tracks are overlaid with color coding for the corresponding [Fe/H] of the tracks, and the red dashed line describes our selection condition for CHeB candidates (see Sect.~\ref{sec:candi}). 
{{\it Left}: We superposed all points of evolutionary tracks with the following conditions: stars younger than 2~Gyr, with  [Fe/H] ranging from $-2.5$ to $-0.8$, and with masses between 1.8 and 3.0 $M_{\odot}$. \it Right}: We superposed all points of the evolutionary tracks with the following conditions: stars older than 7~Gyr, with [Fe/H] ranging from $-2.5$ to $-0.8$, and with masses above 0.1~$M_{\odot}$. 
Similarly, we indicate a limit of 2-Gyr by the right most part of the dashed line, i.e., when $G_{BP}-G_{RP}>0.9$.
}
\label{fig:selections}
\end{figure*}

\begin{figure*}
\centering
\resizebox{16cm}{!}{\includegraphics{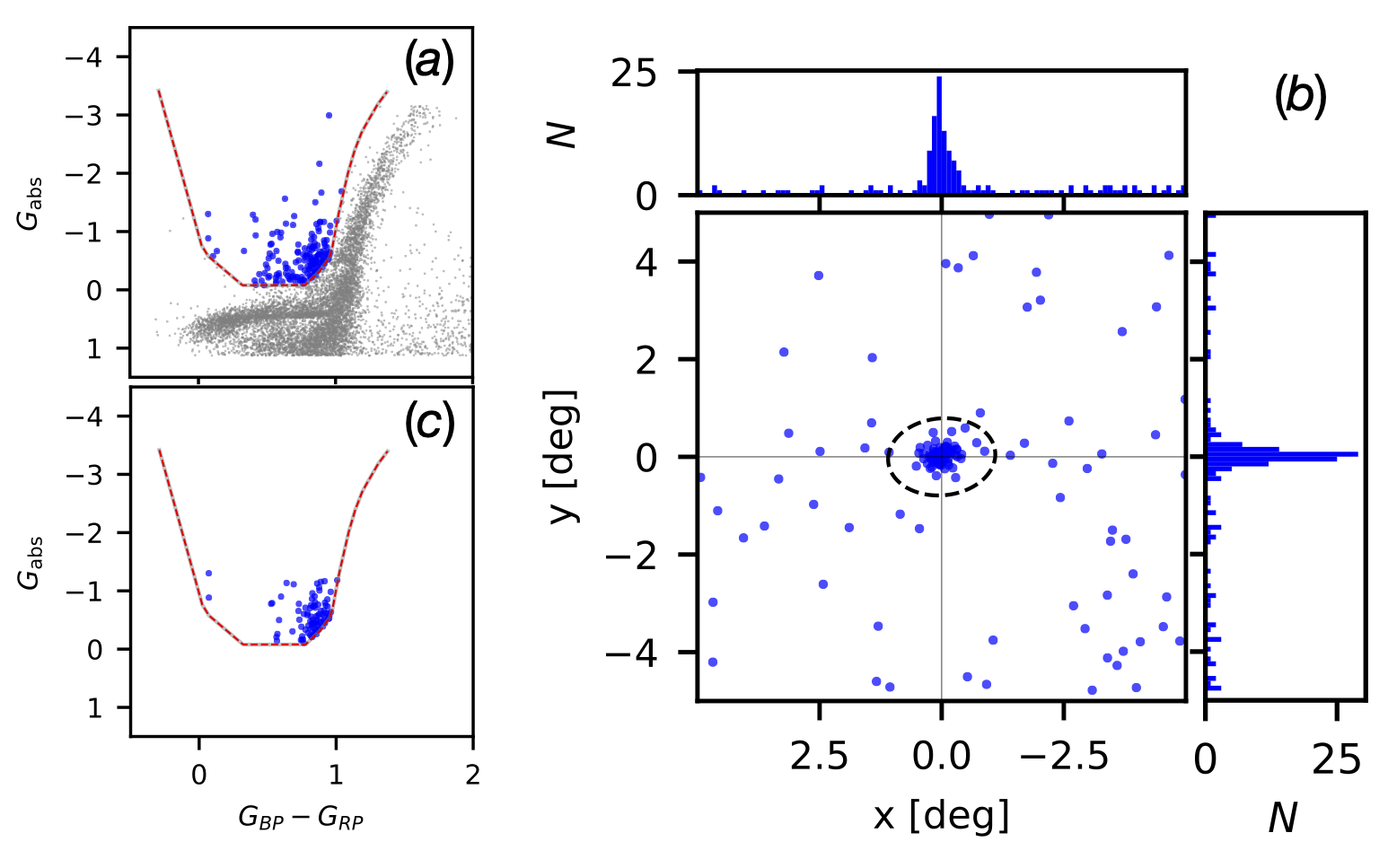}}
\caption{Selection of CHeB candidates and their spatial distribution. Panel-(a), a rough CMD selection of CHeB candidates (blue dots) in the full field of $5\times5$~sq.~deg. The PM-selected sample in the same field is plotted in gray dots. Panel (b), the spatial distribution of the rough CHeB candidates, with histograms of counts projection on each axis for showing the significant concentration of the CHeB candidates at the Sculptor's center. Panel (c) shows only the CHeB candidates that are selected by their locations in the CMD, spatial location within the elliptical radius of 1.1 degree.
The red dashed line in both panel (a) and (c) is the CHeB selection condition (see Figure~\ref{fig:selections} and Sect.~\ref{sec:candi}).
}
\label{fig:sculstat}
\end{figure*}

\begin{table*}
\caption{Statistics of CHeBs in the MW dSphs
}
\begin{center}
\begin{tabular}{lccccccccccc}
\hline\hline
    & (1) & (2) & (3) & (4) & (5) & (6) & (7) & (8)& (9) & (10)  \\
  Name  &  $R_{\rm ell}$ & ($R_{1},R_{2}$) &$N_{\rm r}$ &$N_{\rm net}$  & S/N & $\Sigma_{\rm bg}$ & $N_{\rm Contam.}$ & $p_{\rm MW}$ & $N$&  $N$ \\
      & (deg; kpc)&(deg) & & & & (deg$^{-2}$) & & & ($>1.8\,\msun$) & ($>2.04\,\msun$)   \\ \hline
  Sculptor  & 1.1; 1.65 & (1.2, 5.0) &    91 & 90    & 9.4  & $0.47 \pm 0.07$ & $ 1.3\pm 0.2$ & $5\times10^{-129}$  & 63  & 39 \\ 
   Sextans  & 0.8; 1.18 & (1.0, 5.0) &    54 & 48  & 6.53 & $3.4 \pm 0.19$ & $ 6.0\pm 0.3$ & $1.4\times10^{-32}$  & 37  & 26 \\ 
       UMi  & 1.48; 1.95 & (3.0, 5.0) &    44 & 42  & 6.31 & $0.67 \pm 0.10$ & $ 2.1\pm 0.3$ & $6\times10^{-42}$  & 29  & 22 \\
     Draco  & 0.5; 0.72 & (0.8, 5.0) &    32 & 31  & 5.43 & $2.37 \pm 0.16$  & $ 1.3\pm 0.1$ & $4.6\times10^{-33}$  & 14  & 8  \\
    Carina$^{\dag}$  & 0.4; 0.74 & (0.35, 0.7) &  95 & 85  & 8.63 & $38.7 \pm 4.96$ & $10.2\pm 1.3$ & $2.4\times10^{-57}$ & 69 & 42  \\
    Fornax$^{\dag}$  & 1.3; 3.16 & (1.5, 2.5) &   964 & 923  & 29.6 & $11.3 \pm 0.8$ & $40.8\pm 2.9$ & $1.4\times10^{-933}$ & 687 & 353 \\ 
\hline
\end{tabular} 
\end{center}

\parbox{\hsize}{Notes: Columns~(1-6): Relevant parameters for the selection of CHeB candidates, see Sect.~\ref{sec:candi}.
Column~(7): Contamination counts within the elliptical area covered by $R_{\rm ell}$, inferred from the contamination density $\Sigma_{\rm bg}$.
Column~(8): Probability that all CHeB candidates are due to contamination by MW stars.
Columns~(9-11): Counts of CHeB candidates when considering only stars with mass larger than 1.8 and 2.04~M$_{\odot}$, which correspond to twice the turnoff mass for a stellar population with an age of $\le$ 7 and $\le$ 5 Gyr, respectively.
The parameters for other dSphs than Sculptor are derived from Appendix~\ref{sec:otherdSphs}.\\
$^\dag$ The methodology of Appendix~\ref{sec:otherdSphs} was used to derive parameters for Carina and Fornax, which distances are too large and requires a deeper photometry than that of {\it Gaia}.}
\label{tab:CHeBstat}
\end{table*}

\subsection{CHeB candidates in dSphs}
\label{sec:CHeBselection}
In the following we will concentrate on the analysis of CHeB stars lying in the Sculptor dSph, although the same methodology applies to other dSphs. Results for the latter galaxies are presented in Appendices~\ref{sec:otherdSphs} and \ref{sec:agealgo}.

\subsubsection{Step 1: The selection of CHeB population.}
\label{sec:candi}

Figure~\ref{fig:selections} shows the CMD of Sculptor and compares it to PARSEC evolutionary tracks. 
The right panel identifies stars older than 7 Gyr on the basis of evolutionary tracks for stellar mass larger than 0.1~$M_{\odot}$, and within the Sculptor's metallicity range in Table~\ref{tab:par}. The limit of 7 Gyr corresponds to the dominant stellar population according to \citet{DeBoer12}. 
Similarly, we have indicated a 2-Gyr limit for RGB stars, which is indicated by the red dashed line. 
We may notice that Sculptor may have some stars formed in between the past 2 to 7 Gyr, although these stars would be more difficult to 
identify without ambiguity due to photometric errors on colours and the limited
colour spread of RGBs that encompass this range of ages. 
In the left panel of Figure~\ref{fig:selections}, we superposed all evolutionary tracks for stellar ages smaller than 2 Gyr, within the Sculptor's metallicity range (Table~\ref{tab:par}), and within a mass range of [1.8 to 3.0]~$M_{\odot}$. 
Stars bluer than the 2-Gyr limit and well above the HB are consistent with  intermediate-mass stars in the CHeB phase. 
Of course, we could argue these stars could be also rejuvenated stars that were formed by binaries (i.e., BSSs). We discuss this possibility later (see Sect.~\ref{sec:discussion}), whereas here we pursue our analysis of their properties under the assumption that they are young stars. 

For a more precise selection of the CHeB population, we further exclude AGB, HB (also RR-Lyrae), and MS stars. This leads to the red-dashed line delineated in Figure~\ref{fig:selections}, which defines our method to select CHeB stars, that is to say that selected stars lie above the line. To exclude RR-Lyrae, we set a limit of 0.55 magnitude above the median HB around $G_{BP}-G_{RP} = 0.5$. Since different galaxies have a different magnitude of the  HB, the lower limits of the red-dashed line are different for each galaxy.

It is worth to note that our selection may include young stars in the SG phase, which are also of interest for the goals of our paper. The impact of considering the later stars is small because the SG phase for intermediate-mass stars is relatively short (by a factor of about 6 as shown in Figure~\ref{phases}, for example) when compared to the duration of the CHeB phase. 

\subsubsection{Step 2: Statistical significance of CHeB candidates} 
\label{sect:step2}
Figure~\ref{fig:sculstat} presents our selection criteria of the CHeB candidates in Sculptor. Panel ($a$) shows the CMD of the PM-selected sample (gray dots) within $5\times5$~sq~deg region. We mark in blue the stars that are  consistent with a CHeB phase.
This is a rough selection because it picks up CHeB candidates over an extremely large sky area. 
The stars selected far from  Sculptor may be due to contamination by MW stars.
Panel ($b$) shows the spatial distribution of the rough selection of CHeB stars (see blue dots), which present a strong concentration in the Sculptor center, as also indicated by the peaks of the two histograms of star counts projected on each axis, respectively. 
Besides the concentration in the central region, part of the candidates in the rough selection spread almost homogeneously far from Sculptor over the full field of view, as expected for MW stars. We can consider the latter to evaluate the background level of the contamination by MW stars.

We first calculated the surface density of the contamination background ($\Sigma_{\rm bg}$), by counting the number of the rough CHeB candidates within an annulus from $R_1$ to $R_2$; namely, 1.2 to 5.0 degree from the Sculptor center (see Table~\ref{tab:CHeBstat}). We then derived the MW contamination level, which is found to be $\Sigma_{\rm bg} = 0.47 \pm 0.07$ stars/deg$^2$. This low contamination density level is partly related to the location of Sculptor at high Galactic latitude of $-78.11$~degree, far from the MW disk. 
It also results from the efficient filtering by proper motion, as most of MW stars have distinguishable proper motions when compared to those of Sculptor. 

At this point, we should be able to evaluate the significance of the concentration of the CHeB candidates and their probability of belonging to Sculptor. We define the signal-to-noise ratio (S/N) of the net number count of CHeB candidates using a set of Eqs~(\ref{eq:snr}):
{
\begin{align}
    A &= \pi (1-e) R_{\rm ell}^2, \nonumber \\
        N_{\rm net} &= N_{\rm r} -  A \Sigma_{\rm bg} \ , \label{eq:snr} \\
    {\rm  S/N} &=  N_{\rm net} / \sqrt{N_{\rm r}}\ , \nonumber
\end{align} 
}
where $R_{\rm ell}$ is the elliptical radius (see the dashed-line ellipse in  the panel (b) of Figure~\ref{fig:sculstat}) that describes the morphology of Sculptor together with PA and $e$ (see Table~1), $N_{\rm r}$ gives the counts inside the elliptical area $A$, $N_{\rm net}$ gives the net count after subtracting the MW contamination, namely, $N_{\rm Contam.}$, which is calculated through $N_{\rm Contam.}$=$A \Sigma_{\rm bg}$.
We have let  $R_{\rm ell}$ vary and its adopted value is given after maximizing the signal-to-noise ratio (S/N). 
We have detected a significant number of CHeB candidates ($N_{r}$) in Sculptor, which is far more superior than the expected contamination from the MW ($N_{\rm Contam.}$). We then calculate the probability that MW stars could be mistaken as CHeB candidates, assuming a Poisson distribution for counts. The probability of $k$ events for a mean count of $\lambda$ is given by:
$p_k = {\lambda ^{k}e^{-\lambda }}/{k!}$.
For Sculptor, it leads to a  count of 91 CHeB stars within $r_{\rm ell} < 1.1$~deg. The MW contamination in the same area is $1.3 \pm 0.2$ stars. Then the probability of 91 counts come the fluctuation of 1.3 is $p_{\rm MW} = 5\times10^{-129}$, which rules out the possibility of a contamination by MW stars.
All the above quantities are listed in Table~\ref{tab:CHeBstat}, together with those for other dSphs (see also Appendix~\ref{sec:otherdSphs}).

\subsection{Ages, masses, and metallicities  of CHeB stars in dSphs}
\label{allages}

\begin{table*}
\caption{\bf Age and mass estimates for the Sculptor  stars observed with Giraffe. }
\label{ages}
\tabcolsep=4pt
\begin{center}
\begin{tabular*}{\textwidth}{@{\extracolsep{\fill}}lrrrrrrrrrcrrrrrr}
\hline\hline
 & \multicolumn{9}{c}{Method tracks}& & \multicolumn{6}{c}{Method isochrones}\\
 \cline{2-4}
 \cline{5-7}
 \cline{8-10}
 \cline{12-14}
 \cline{15-17}
 & \multicolumn{3}{c}{solution0} & \multicolumn{3}{c}{solution1} & \multicolumn{3}{c}{solution2} 
 &  &
 \multicolumn{3}{c}{solution1} & \multicolumn{3}{c}{solution2}\\ \hline
Name &
  \multicolumn{1}{c}{age} &
  \multicolumn{1}{c}{mass } &
  \multicolumn{1}{c}{phase } &
  \multicolumn{1}{c}{age} &
  \multicolumn{1}{c}{mass } &
  \multicolumn{1}{c}{phase } &
  \multicolumn{1}{c}{age}  &
  \multicolumn{1}{c}{mass}  &
  \multicolumn{1}{c}{phase}  & 
 &  \multicolumn{1}{c}{age} 
  & 
  \multicolumn{1}{c}{mass} &
  \multicolumn{1}{c}{phase } &
  \multicolumn{1}{c}{age}  &
  \multicolumn{1}{c}{mass}  &
  \multicolumn{1}{c}{phase}  \\
 & \multicolumn{1}{c}{(Myr)}   & \multicolumn{1}{c}{(M$_\odot$)}  & \multicolumn{1}{c}{} & 
 \multicolumn{1}{c}{(Myr)}   & \multicolumn{1}{c}{(M$_\odot$)}  & \multicolumn{1}{c}{} & 
 \multicolumn{1}{c}{(Myr)}   & \multicolumn{1}{c}{(M$_\odot$)}  & \multicolumn{1}{c}{} 
 & &
 \multicolumn{1}{c}{(Myr)}   & \multicolumn{1}{c}{(M$_\odot$)}  & \multicolumn{1}{c}{} & 
 \multicolumn{1}{c}{(Myr)}   & \multicolumn{1}{c}{(M$_\odot$)}  & \multicolumn{1}{c}{}  \\
\hline
  Scl2    &    629 & 2.0 & CHeB  &    629 & 2.0 &  CHeB  &   473 & 2.4  & RGB    & & 499 & 2.5 & CHeB & 389 & 2.7 & SG\\
  Scl3    &    494 & 2.4 & CHeB  &    494 & 2.4 &  CHeB  &   494 & 2.4  & CHeB   & & 722 & 2.2 & CHeB & 331 & 2.8 & SG \\
  Scl8    &    544 & 2.2 & CHeB  &   1206 & 1.8 &  AGB   &   512 & 2.4  & CHeB   & & 731 & 2.2 & CHeB & 358 & 2.8 & SG \\
  Scl9    &    467 & 2.4 & CHeB  &    790 & 2.0 &  CHeB  &   467 & 2.4  & CHeB   & & 703 & 2.3 & CHeB & 317 & 2.8 & SG\\
  Scl12   &    912 & 1.9 & CHeB  &    912 & 1.9 &  CHeB  &   848 & 1.9  & AGB    & & 757 & 2.1 & SG                      & 339 & 2.8 & SG\\
  Scl19   &    781 & 2.1 & CHeB  &   4715 & 1.1 &  RGB   &   468 & 2.4  & RGB    & & 399 & 2.8 & SG                      &     &     &  \\
  Scl29   &    436 & 2.4 & CHeB  &    962 & 1.9 &  AGB   &   436 & 2.4  & CHeB   & & 795 & 2.0 & AGB & 752 & 2.1 & CHeB \\
  Scl31   &    980 & 1.9 & CHeB  &   1301 & 1.7 &  CHeB  &   581 & 2.3  & CHeB   & & 1356& 1.7 & CHeB & 482 & 2.3 & SG\\
  Scl38   &   1016 & 1.9 & CHeB  &   1171 & 1.6 &  CHeB  &   537 & 2.4  & CHeB   & & 1393& 1.7 & CHeB & 521 & 2.3 & RGB \\
  Scl47   &    608 & 2.1 & CHeB  &   1403 & 1.6 &  RGB   &   521 & 2.3  & RGB    & & 608 & 2.1 & CHeB                    & 555 & 2.2 & RGB \\
  Scl61   &    782 & 2.1 & CHeB  &    852 & 2.0 &  CHeB  &   613 & 2.2  & CHeB   & & 2168& 1.4 & CHeB & 658 & 2.0 & CHeB\\
  Scl68   &    655 & 2.2 & CHeB  &   1062 & 1.8 &  CHeB  &   523 & 2.4  & CHeB   & & 962 & 1.9 & CHeB & 394 & 2.6 & SG\\
  Scl70   &    723 & 2.2 & CHeB  &    957 & 1.9 &  CHeB  &   488 & 2.3  & CHeB   & & 904 & 2.0 & CHeB & 364 & 2.6 & SG\\
  Scl75   &    740 & 2.1 & CHeB  &   1309 & 1.7 &  CHeB  &   630 & 2.2  & CHeB   & & 1144 & 1.8 & CHeB & 449 & 2.4 & SG\\
 \hline
 \end{tabular*}
 \end{center}

 \parbox{\hsize}{Two types of solutions from ``method tracks" and ``method isochrones" are provided. The solutions ``1" and ``2" correspond to the lowest and highest possible mass in different methods, respectively. The solution "0" of the "method tracks" is the most probable solution based on the elapsed-time argument.}
\end{table*}

By definition, our selection of CHeB stars is presumably made of young stars, namely, with ages smaller than 2-Gyr. Based on their locations in the CMD, we further investigated the details of their stellar parameters, such as masses, metallicities, stellar ages, 
and evolutionary  phases for the CHeB candidates. 
For five out of the six galaxies studied, we only have photometric data
to perform this estimate. For Sculptor, instead, we also have spectra, that 
allow to derive metallicities, this breaks the age-metallicity degeneracy, that
is one of the sources of uncertainty.
We tested two methods, one (i.e., "method tracks") based on evolutionary tracks, and no information
on metallicity for individual stars beyond what can be inferred from CMD. 
The other one (namely "method isochrones") is based
on isochrones, which we applied only to Sculptor, which makes use of the
spectroscopic metallicities discussed in Sect.\,\ref{spec}.
In principle, the two methods should provide consistent results, because both methods are based on the same evolutionary track library, PARSEC. In practice, the way of building and using tracks and isochrones (e.g., parameter spacing in age, mass, and metallicities, or algorithms used to pick a solution) may cause some differences in results.
The results are provided in Table\,\ref{ages} and the main conclusion
is that whatever the details of the method used to estimate ages and masses, 
the ages of these stars are young and their masses large
enough to classify them as intermediate mass stars. 
That is stars that, when on the main sequence were of spectral type A or B.

\begin{figure}
\centering
\resizebox{8cm}{!}{\includegraphics{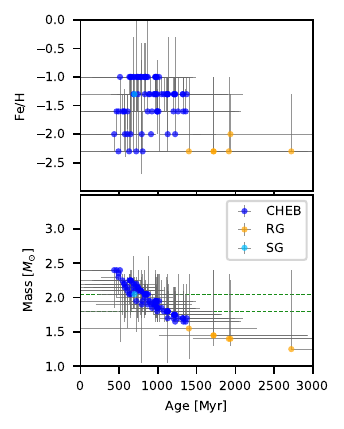}}
\caption{Age-metallicity (left), and age-mass (right) relations for CHeB stars in their different evolutionary phases (indicated by color explained in the legend) for Sculptor. The three dashed lines in the middle panel indicate the masses of 1.8 and 2.04~M$_{\odot}$, respectively. 
}
\label{CHeBages}
\end{figure}

\subsubsection{Ages, masses, and metallicities based on photometry and evolutionary tracks}
\label{agenospec}

We first prepared a sample of evolutionary tracks within the metallicity range of each dSph (see Table~\ref{tab:par}) and within a mass range of [1, 8]~$M_{\odot}$, respectively. 
The mass range is wide enough to cover all CHeB candidates. 
We have let  the stellar age of the tracks free in this calculation to avoid possible cutoff biases.

For a given star, the mass, metallicity, age, and evolution phase have been estimated through all possible evolutionary tracks that are covered by 1 sigma error in photometry. To identify the most probable evolutionary track, we defined a parameter of elapsed time that a star will spend to pass on a given track, after accounting for the window of the error bar. Therefore, the slow evolutionary track is more likely to be observed. Appendix~\ref{sec:agealgo} describes  the algorithms  we used for the solutions reported for individual stars.
The final results for CHeB candidates of Sculptor is presented in Figure~\ref{CHeBages}, and Figure~\ref{CHeBagesA} for other dSphs.

\subsubsection{Ages and masses based on photometry, isochrones, and spectroscopic metallicities, in Sculptor}
\label{agewithspec}

For the 14 stars in Sculptor discussed in Sect.\,\ref{spec}, we can take
advantage of the spectroscopic metallicities, thus breaking the age-metallicity degeneracy.
To estimate ages and masses, we use 
the color magnitude diagram $G_{BP}-G_{RP}$ vs. $G_{\rm abs}$. 
We compared each star to 
PARSEC \citep{bressan12} isochrones, within 0.3\,dex of the
spectroscopic metallicity, corresponding to 7 metallicity bins.
The age steps were of 100\,Myr.
We then selected among the isochrones  the same evolutionary stage. 
The complex topology of the isochrones, as illustrated in Fig.\,\ref{phases}
makes this exercise ambiguous.
For each star, for each age and evolutionary phase, 
we interpolated in the isochrone the $G_{\rm abs}$ corresponding to
the star's $G_{BP}-G_{RP}$. Then on this curve in the CMD,  we interpolated
the values of age and mass corresponding to the observed $G_{\rm abs}$.
For most evolutionary phases, there was no solution, since $G_{\rm abs}$ was either
larger than the maximum or smaller than the minimum values on the curve.
We derived two or three solutions for each star, only for star
Scl19 we could find only one solution.
In Table\,\ref{ages}, we report only two solutions, corresponding to the lowest
and highest possible mass. For each star we also report
the corresponding evolutionary stage.

To verify the impact of distance error to the stellar parameters estimates, we repeated 
the above estimation procedures by considering an additional error in the magnitude of stars due to the uncertainty of distance modulus (Table~\ref{tab:par}). We found that the differences of the two estimations show an unbiased scatter of 24\% in age and 8\% in mass determination, respectively. These scatters are small compared to the uncertainty due to the complexity of stellar evolution, which is usually larger than 50\%-100\% in age and about 20\% in mass, as shown in Figure~\ref{CHeBages}. 
We also note that the counts in columns 9 and 10 in Table~\ref{tab:CHeBstat} are marginally affected within the Poisson statistical errors when considering the error in distance, specifically, $N(>1.8\,\msun)=69$ and $N(>2.04\,\msun)=35$ for Sculptor.
Thus,
the uncertainties of 5\,kpc \citep{DeBoer12} in the distance of Sculptor provides an uncertainty 
in mass and age negligible with respect the uncertainty 
related to the  phase in which the stars are. We applied the same test to the other dSphs in Table~\ref{tab:CHeBstat} using the "method tracks" and our conclusion on the impact of distance error stands the same as for Sculptor.

\section{Spectroscopic study of the CHeB population in Sculptor}
\label{spec}

\subsection{GIRAFFE spectra}
\label{dataspec}

\begin{figure}
    \centering
    \resizebox{8.5cm}{!}{\includegraphics[clip=true]{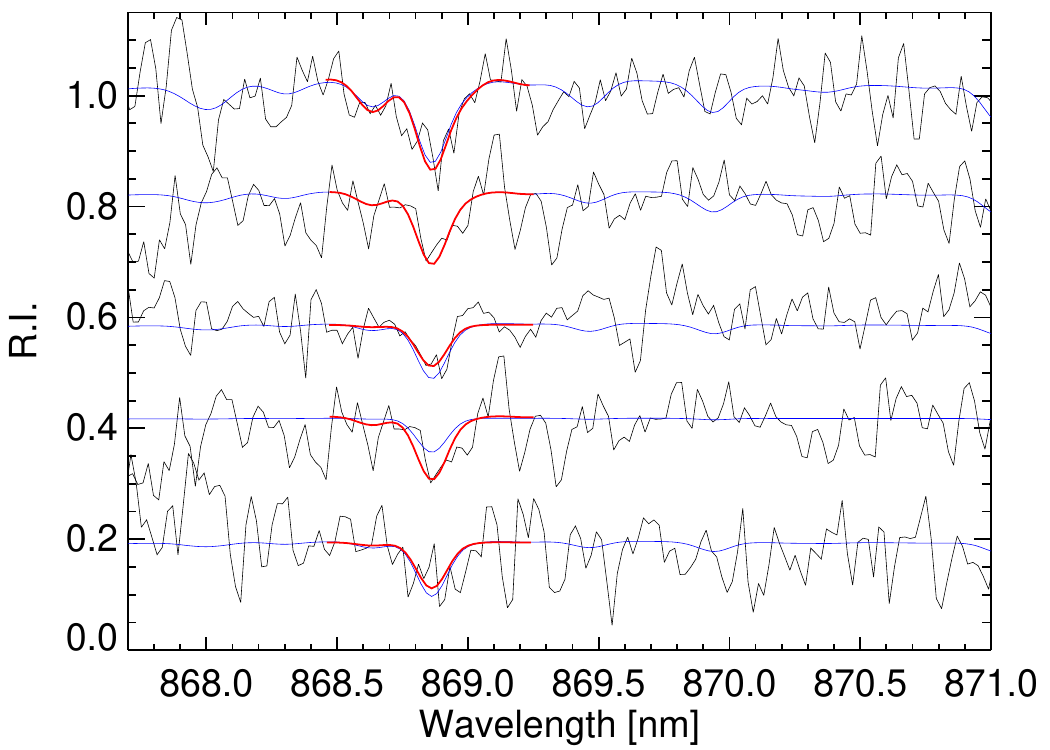}}
    \caption{Observed spectra (solid black) normalised and vertically shifted for presentation purpose, compared to synthetic spectra (solid blue) in the wavelength range of the 868.86\,nm \ion{Fe}{i} line and the best fit of the Fe feature (solid red).
The spectra are ordered in decreasing metallicity, from top (the most metal-rich) to bottom (the most metal-poor): Scl\,12, Scl\,19, Scl\,9, Scl\,29, and Scl\,2.}
    \label{fig:spectra}
\end{figure}

We cross-matched our CHeB candidates (see Sect.~\ref{sec:candi}) with the ESO archive. 
A sample of 28 Giraffe spectra (setting LR08 for detecting the \ion{Ca}{ii} IR triplet) 
of CHeB star candidates was  selected satisfying the CMD selection shown in Fig.\,\ref{fig:sculstat},
with the additional   condition that S/N $\ge 20$ (from the ESO archive). The targets are listed in Table~\ref{tab:vrad}.
The selected spectra belong to programs
0100.B-0337, 0101.B-0189 (P.I. E. Tosltoy), and 0101.D-0210
(P.I. A. Sk{\'u}lad{\'o}ttir).
All the spectra we retrieved from the ESO archive
were previously   analyzed and discussed by \citet{tolstoy23}. Here, we
provide an independent analysis and discuss the similarities and differences with \citet{tolstoy23}.
The ESO archive provides extracted spectra, correction for the earth's motion
(the heliocentric correction), but the spectra are not sky-subtracted.
A median sky spectrum was  computed from the sky spectra accompanying each spectrum from the ESO archive and then subtracted from the object spectrum.
A spectral interval around one of the \ion{Fe}{i} lines used is shown in Fig.\,\ref{fig:spectra}, along with the relevant synthetic spectra.
The radial velocities (see Sec.\ref{radvel}) and abundances (see Sect.\,\ref{atmabund}) derived from the spectra allowed us to confirm or reject the membership of the stars
to the galaxy.

\subsection{Radial velocities}
\label{radvel}

For each star, we measured the radial velocity using our own code for template matching.
The template consists of a synthetic stellar spectrum
chosen according to our estimate for the atmospheric parameters
and a telluric absorption theoretical spectrum from \citet{2014A&A...564A..46B}.
The stellar instrumental radial velocity and that of the telluric lines
were adjusted independently, and the final radial velocity was obtained
by subtracting the velocity of the telluric lines from the instrumental
radial velocity. 
The measured radial velocities are given in Table~\ref{tab:vrad}.
For one star in Table~\ref{tab:vrad}, we have 
two spectra (Scl21 and Scl21a) from program 0101.B-0189 and these were treated independently.  There is a difference of 25.5\,\kms\ between the two measurements. 
The two spectra were observed at different epochs, separated by approximately one year.
The simplest explanation is that the star Scl21 is a binary; thus, we excluded it from the chemical analysis.
\citet{battaglia08apj} measured a mean radial velocity of 110.6\,\kms\  with a dispersion of
10.1\kms for Sculptor, on the basis of 470 stars having Giraffe spectra.
We considered all stars having a radial velocity that lies within three times the velocity dispersion 
from the mean
radial velocity of \citet{battaglia08apj} as Sculptor members. 
This  criterium excludes targets  Scl0 and Scl50,  leaving us with 25 radial velocity members in the sample, and then 24 after removing Scl21.
The mean radial velocity of our retained 24 stars is 109.6 \kms\ with a standard
deviation of 9.8\,\kms\ in excellent agreement with the value of 10.1\,\kms\ derived by  \citet{battaglia08apj}. 
The spectra we analyzed were also previously studied by \citet{tolstoy23}.
For the 24 member stars in our sample (which were also considered as Sculptor members in that study),
their mean radial velocity is 109.6\,\kms\ with a standard deviation of 9.1\,\kms\ ; this is in
excellent agreement with our result. The average difference, in the sense our radial velocities minus
\citet{tolstoy23}, is $-0.02$\,\kms\ with a dispersion of 5.9\,\kms. 
Using 3-$\sigma$ criteria, we excluded 2 out of the 27 stars (including a binary) listed in Table~\ref{tab:vrad} from membership. This suggests that the PM-selected sample is effective, with 25 out of 27 stars identified as potential members, even though the size of this verification sample is statistically limited.

\begin{table} 
\centering
\caption{Radial velocities of the analyzed stars}
\label{tab:vrad}
\begin{tabular}{lccc}
\hline\hline
  Target & RA &  Dec  & $v_{\rm rad}$  \\
  & (deg J2000) & (deg J2000) & (\kms) \\
\hline
Scl0 & $15.549990$ & $-33.718388$ & $ \ \,78 \pm   4$ \\  
Scl1 & $14.995166$ & $-33.714305$ & $118 \pm   6$   \\
Scl2 & $15.525830$ & $-33.662277$ &  $104 \pm   1$  \\
Scl3 & $15.302080$ & $-33.758472$ &  $110 \pm   1$      \\
Scl5 & $15.404990$ & $-33.783555$ &  $104 \pm   7$      \\
Scl8 & $15.055830$ & $-33.754583$ &  $106 \pm   6$      \\
Scl9 & $14.939249$ & $-33.688861$ &  $130 \pm   1$  \\
Scl12 & $15.696240$ & $-33.778249$ & $ \ \,93 \pm   3$ \\
Scl16 & $15.140749$ & $-33.775027$ & $ 106 \pm   1$ \\
Scl18 & $15.159208$ & $-33.808833$ & $ 104 \pm   2$ \\ 
Scl19 & $15.219833$ & $-33.690944$ & $ 112 \pm   4$     \\
Scl21 & $15.458740$ & $-33.845388$ &  $127 \pm   3$ \\ 
Scl21a & $15.458740$ & $-33.845388$ & $102 \pm   2$ \\ 
Scl23 & $15.181958$ & $-33.717388$ & $112 \pm   1$  \\
Scl24 & $14.831166$ & $-33.627361$ & $121 \pm   1$  \\
Scl27 & $14.893208$ & $-33.511333$ & $112 \pm   1$      \\
Scl29 & $15.138708$ & $-33.620277$ & $112 \pm   1$ \\ 
Scl31 & $15.158166$ & $-33.622749$ & $109 \pm   1$ \\ 
Scl38 & $15.372083$ & $-33.851805$ & $124 \pm   3$ \\
Scl40 & $15.268166$ & $-33.688972$ & $111 \pm   1$ \\     
Scl47 & $14.770208$ & $-33.631972$ & $112 \pm   1$ \\      
Scl50 & $14.943333$ & $-33.960083$ & $146 \pm   1$ \\  
Scl59 & $14.745166$ & $-33.940749$ & $108 \pm   1$ \\ 
Scl61 & $15.466249$ & $-33.648722$ & $\ \,83 \pm   1$ \\ 
Scl68 & $14.637208$ & $-33.710361$ &  $\ \,99 \pm   1$ \\
Scl69 & $14.648083$ & $-33.556722$ &  $103 \pm   3$ \\  
Scl70 & $14.699624$ & $-33.495861$ & $117 \pm   3$ \\     
Scl75 & $15.152041$ & $-34.101361$ & $121 \pm   1$ \\ 
\hline\end{tabular}
\end{table}

\begin{figure}
\centering
\hspace{0.3cm}\resizebox{7.5cm}{!}{\includegraphics{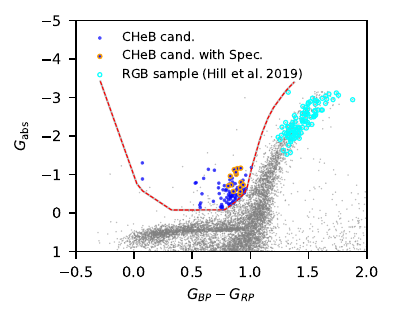}} 
\resizebox{8cm}{!}{\includegraphics{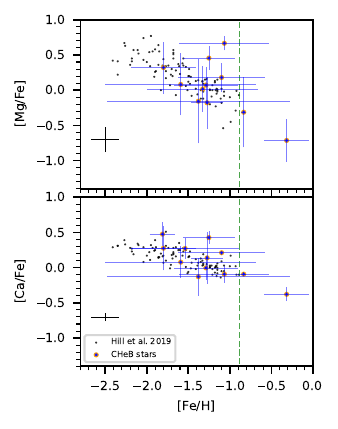}} 
\caption{
{\it }  CMD of Sculptor showing different samples (top). Blue dots for the CHeB candidates; blue dots with orange circle for the CHeB candidates with Giraffe spectra with S/N $\ge 20$; open cyan circles for the RGB sample from \citet{hill19}. The gray dots and the red dashed line have the same meanings as in Figure~\ref{fig:sculstat}. 
Middle and bottom panels give a comparison of the measured metallicities for the CHeB candidates with those of RGB stars measured by \citet[][small black dots]{hill19}. The typical error bars of different elements for the RGB stars are plotted in black on the lower-left of each panel, respectively. In these panels, the green dashed line indicates the upper limit of [Fe/H] for the RGB sample.
}
\label{mgfe}
\end{figure}

\begin{table*}
\caption{Atmospheric parameters and chemical abundances of the stars in this study.}
\label{abund}

\setlength\tabcolsep{3pt}
\centering
\begin{tabular}{lccccccccccccccc}
\hline\hline
  \multicolumn{1}{c}{Target} &
  \multicolumn{1}{c}{$T_{\rm eff}$} &
  \multicolumn{1}{c}{$\log g$} &
  \multicolumn{1}{c}{$v_{\rm vturb}$} &
  \multicolumn{1}{c}{[Fe/H]} &
  \multicolumn{1}{c}{$\sigma(\rm Fe I)$} &
  \multicolumn{1}{c}{$N$(FeI)}&
  \multicolumn{1}{c}{[Mg/H]} &
  \multicolumn{1}{c}{[Mg/Fe]} &
  \multicolumn{1}{c}{$\sigma(\rm Mg I)$} &
  \multicolumn{1}{c}{$N$(MgI)} &
  \multicolumn{1}{c}{[Ca/H]} &
  \multicolumn{1}{c}{[Ca/Fe]} &
  \multicolumn{1}{c}{$\sigma(\rm  CaII)$} &
  \multicolumn{1}{c}{$N$(CaII)} \\
        & (K) &  & (\kms)\\
\hline
Scl2    &5354 &2.56   &1.66   &$-1.539$ &0.650  &1      &$      $  &$      $ &      &0       &$-1.267$  &$0.272 $&0.050     &1 \\
Scl3    &5511 &2.45   &1.77   &$-1.066$ &0.543  &2      &$-0.400$  &$0.666 $ &0.100 &1       &$-1.157$  &$-0.091$&0.040     &1 \\
Scl8    &5380 &2.54   &1.64   &$-0.833$ &0.027  &3      &$-1.146$  &$-0.313$ &0.500 &1       &$-0.925$  &$-0.092$&0.050     &1 \\
Scl9    &5429 &2.42   &1.75   &$-1.099$ &0.522  &3      &$-0.919$  &$0.180 $ &0.200 &1       &$-0.886$  &$0.212 $&0.010     &2 \\
Scl12   &5551 &2.50   &1.71   &$-0.316$ &0.269  &3      &$-1.031$  &$-0.714$ &0.300 &1       &$-0.694$  &$-0.378$&0.033     &1 \\
Scl19   &5355 &2.61   &1.63   &$-1.092$ &0.753  &2      &$      $  &$      $ &      &0       &$      $  &$      $&          &0 \\
Scl29   &5327 &2.24   &1.82   &$-1.326$ &0.669  &2      &$-1.308$  &$0.017 $ &0.320 &1       &$      $  &$      $&          &0 \\
Scl31   &5621 &2.56   &1.84   &$-1.811$ &0.155  &2      &$      $  &$      $ &      &0       &$-1.336$  &$0.476 $&0.057     &2 \\
Scl38   &5368 &2.56   &1.65   &$-1.252$ &0.309  &2      &$-0.796$  &$0.456 $ &0.170 &1       &$-0.826$  &$0.427 $&0.030     &1 \\
Scl47   &5258 &2.38   &1.74   &$-1.799$ &0.396  &4      &$-1.478$  &$0.321 $ &0.370 &1       &$-1.522$  &$0.276 $&0.105     &2 \\
Scl61   &5499 &2.53   &1.77   &$-1.591$ &0.907  &2      &$-1.510$  &$0.081 $ &0.440 &1       &$-1.514$  &$0.078 $&0.076     &3 \\
Scl68   &5594 &2.52   &1.80   &$-1.274$ &0.191  &2      &$-1.453$  &$-0.179$ &0.370 &1       &$-1.134$  &$0.139 $&0.123     &2 \\
Scl70   &5444 &2.45   &1.76   &$-1.288$ &0.381  &3      &$-1.215$  &$0.073 $ &0.260 &1       &$-1.291$  &$-0.003$&0.056     &2 \\
Scl75   &5574 &2.56   &1.77   &$-1.379$ &1.098  &2      &$-1.536$  &$-0.157$ &0.600 &1       &$-1.508$  &$-0.129$&0.090     &1 \\
\hline
\end{tabular}
\parbox{\hsize}{
Notes: $N$(FeI), $N$(MgI), and $N$(CaII) are the numbers of FeI, MgI, and CaII lines, respectively, in the spectra of each star.}
\end{table*}



\subsection{Atmospheric parameters and abundances}
\label{atmabund}

The atmospheric parameters were derived from the {\it Gaia} photometry,
as described in \citet{lombardo21}, namely, assuming 
an $E(B-V)$ and a distance in Table~\ref{tab:par} for Sculptor.
For each star, we assumed the highest mass listed in Table\,\ref{ages},
which impacts the derived surface gravity and thus the \ion{Ca}{ii} abundances.
The micro-turbulent velocity was derived from the calibration of 
\citet{mashonkina17}.
We then ran \mygi\ \citep{mygi} to derive the abundances of Fe, Mg, and Ca of CHeB candidate stars.
We were finally left with 14 stars, for which the spectra have sufficient S/N values to provide
an acceptable accuracy for their abundances. 
The atmospheric parameters and abundances are given in Table~\ref{abund}.

In Fig.~\ref{mgfe}, we compare our abundances with those of RGB stars from \citet{hill19}. 
By and large, our measurements are compatible with those of  \citet{hill19}. Our errors
are considerably larger, in the first place because  our targets are 3 to 1.5 magnitudes
fainter than those of \citet{hill19}; in the second place, we have a lower resolution and a
smaller spectral coverage.
Two facts are worth noting: our sample extends
to higher metallicities with respect to that of \citet{hill19}
and it also lacks the most metal-poor stars that are present
in the \citet{hill19} sample.
The comparison with the analysis of \citet{tolstoy23} is less straightforward.
While we measured three elements from these spectra, \citet{tolstoy23} measured
the equivalent widths of the \ion{Ca}{ii} 854.2\,nm and 866.2\,nm lines and
used the calibration of \citet[][Equation (5)]{starkenburg10}
to derive the metallicities. A direct comparison of our
[Fe/H] values with those of \citet{tolstoy23} shows a large discrepancy,
with our values being systematically more metal-rich, up to +1.4\,dex.
However, one has to keep in mind that the \citet{starkenburg10}
calibration has been derived and is valid only for RGB stars 
as duly noted by \citet{tolstoy23}, who underline that their metallicities
are valid only for RGB stars.
In fact the effective temperature of the star is implicitly 
taken into account through its $V$ magnitude\footnote{In fact, through the difference
between its $V$ magnitude and that of the Horizontal Branch.}.
An inspection of Fig.\,\ref{mgfe} (top panel) shows that while there is a clear
relation between magnitude and color (and, therefore, \teff ),  along
the RGB, this does not exist among CHeB stars, which occupy
a narrow range in color, but span almost one magnitude in 
$G_{\rm abs}$, with no clear correlation between the two.
We therefore conclude that the discrepancy between our [Fe/H] values
and those of \citet{tolstoy23} is mainly due to the the fact
that the \citet{starkenburg10} calibration is not applicable to CHeB stars.

\section{Discussion} \label{sec:discussion}
In this paper, 
using theoretical evolutionary tracks, we have been able to estimate  masses and ages for
stars that occupy the region expected for young stars in the
CHeB phase. In this way, for Sculptor, we have identified  a population of 91 young and massive stars with ages from 250 to 1500 Myr old, with masses of $\ge$ 1.6 $\rm M_\odot$; this includes 63 (69\%) and 39 (43\%) stars that are more massive than 1.8 and 2.04\,$\rm M_\odot$, respectively (see the last columns of Table~\ref{tab:CHeBstat}). Although some could also be in the SG phase; if so,
that indicates they are even younger and more massive than if they were in the CHeB phase. 
In principle, our selection of CHeB may include the Type II Cepheids which are old ($\sim10$~Gyr) and low-mass stars of about 0.5--0.6~$\msun$\citep{Bono2020}. 
By comparing to the {\it Gaia} Cepheid catalogs \citep{Clementini2023,Ripepi2019}\footnote{This includes a direct cross-matching with the Cepheid table in DR3, namely \texttt{vari\_cepheid}.}, we found none of our CHeB candidates to be a type II Cepheid. 
This suggests that the occurrence of type II Cepheids is very rare in the MW dSphs, which is consistent with former studies \citep[see][]{Monelli2022}.
We need to verify whether alternative explanations, other than recent star formation, 
may be consistent with the observations. In the following we examine two
possibilities:  either CHeB candidates are ``rejuvenated'' stars (i.e., evolved  BSSs) 
or they are truly young and have originated in the  
turbulent star formation events associated to ram pressure caused by a recent entry of dwarfs into the MW halo (see Papers I, II, and III).

\subsection{Interpretation of CHeB candidates in dSphs as evolved BSSs}
\label{sec:BS}
We want to discuss the possibility of interpreting the CHeB stars
as evolved BSSs. 
This requires a possible connection between these stars and
the stars that appear in the blue plume and on the 
upper main sequence. We discuss these two
regions of the CMD separately.

\subsubsection{Connection with the blue plume}
\citet{Momany2007} carried out an extensive study of
blue plumes in Local Group dSphs, excluding Fornax,
known to possess a young population. 
They established  an anticorrelation between 
the fraction of BSSs and the luminosity of the galaxy. They concluded that this anti-correlation could
be used to discriminate blue plumes made up mainly by BSSs,
which follow this anti-correlation, 
and blue plumes that contain a significant young population,
such as Carina, that  do not follow this anticorrelation. 
They never excluded the presence
of a small fraction of young stars, even in blue plumes
dominated by BSSs stars.
A similar study was performed by \citet{santana13}, with
different data and techniques. They found that the 
fraction of BSSs is either independent of the galaxy luminosity, 
or there is a mild anticorrelation, similar to that found
by \citet{Momany2007}.
They also concluded that the blue plumes in the galaxies studied by them
could comprise also a small fraction (1\%--7\%) of young (age of the order of 2.5 Gyr)
stars.

The blue plume stars  in Sculptor and Fornax have been studied in detail
by \citet{Mapelli2009}. Based on considerations on the radial 
distribution of BSSs stars and on dynamical simulations
with the code of \citet{SP1995}, they conclude that while
in Fornax the ``blue plume'' comprises a mixture
of young stars and BSSs, the ``blue plume'' of Sculptor
can be well explained as being made exclusively by BSSs.
However, in their appendix A1, \citet{Mapelli2009} also point out
that the interpretation of the ``blue plume'' as composed
by stars of age 2-3 Gyr cannot be ruled out.

\begin{figure}
\centering
\includegraphics[width=\hsize]{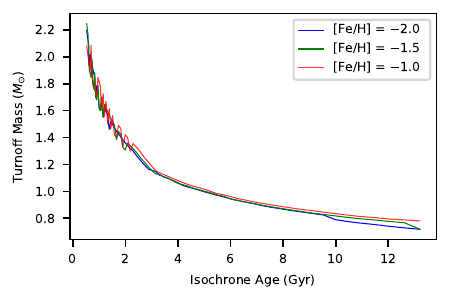}
\caption[]{
Turnoff mass of isochrones of different metallicities and ages.
}
\label{turnoff}
\end{figure}

\citet[see their Fig. 8]{DeBoer12} found that 95.5\% (84\%) of Sculptor dSph stars are older than 8 Gyr (10 Gyr), respectively. 
All these stars are metal poor (Fe/H $< -1.5$).  
This is supported by \citet{Bettinelli2019} who found the whole stellar population in Sculptor to be older than 11 Gyr. For an old ($\ge$ 10 Gyr), and metal poor stellar population, a turnoff star has a mass smaller than 0.8 $\rm M_\odot$ (see Figure~\ref{turnoff}). If we account for the $\sim$ 11.5\% of stellar ages between 8 and 10 Gyr, the turnoff mass limit goes to 0.9 $\rm M_\odot$. We consider these two values as the real limits for producing BSS from binary stars, since to reach 1 $\rm M_\odot$ for the turnoff mass needs to consider the less than 1\% stars having an age of 5 Gyr old according to the SFH from \citet{DeBoer12}.  
A higher mass than the turnoff would imply that the star has already completed its evolution and would have already evolved to the white dwarf stage. If a BSS is formed through mass accretion in a binary system or from a fusion of two stars it should not have a mass larger than  twice that of a turnoff star; namely, less than 1.6, and (for a small fraction) 1.8 $\rm M_\odot$. 
In Figure~\ref{phases}, we show an evolutionary track of a simulated BSS formed by the fusion of two 0.8~$\rm M_\odot$ stars, resulting in a star of 1.57~$\rm M_\odot$.
 However, Table~\ref{tab:CHeBstat} shows that the majority (63 among 91, i.e., 69\%) of CHeB candidates possess masses well above 1.8 $\rm M_\odot$, whichever the way we estimate the mass. 
Blue stragglers that have masses larger than twice the mass of turnoff stars are known both in open clusters \citep[e.g., M\,67,][]{milone92} and globular clusters \citep[e.g., NGC\,6397,][]{saffer02}.
In Appendix \ref{triples}, we estimate how efficiently a fusion
of a triple system can form massive BSSs. We have concluded that this extreme mechanism is unlikely able to explain our massive CHeB population.

If we consider our spectroscopic analysis of Sculptor stars, Fig.\,\ref{mgfe} shows that the CHeB candidates are 
relatively more metal rich, on average, 
with a lack of stars at the low metallicity end and a few stars showing higher
metallicity with respect to the RGB sample of \citet{hill19}. 
This does not favour a BSS origin for most CHeB candidates, 
for which one would have expected a similar distribution of metallicity than for the RGB general population from which they should come.
To summarize, the four difficulties with interpreting CHeB candidates as BSSs in Sculptor dSph are as follow. 
\begin{enumerate}
\item It is not
possible for   a BSS to form with a mass of $\ge 1.8\,\msun$
from an old population of stars 
with masses $\le 0.8-0.9 \msun$ without
involving at least three such stars. It appears unlikely that all these stars were formed in triple or
higher multiplicity systems.
\item We found 39 CHeB stars with mass larger than 2.04 $M_{\odot}$ (see the last column of Table~\ref{tab:CHeBstat}), while none of them can be predicted from the turnoff mass -- even if their progenitors had, in fact, come from the rare ($<$1\%) population of 5 Gyr old stars (turnoff mass of 1 $M_{\odot}$, see Figure~\ref{turnoff}), according to the SFH from \citet{DeBoer12}. 
\item The lack of evolved blue stragglers with [Fe/H] $< -2.0$. 
\item Two stars (out of 14, i.e., 14\%) more metal-rich than the most metal-rich stars observed on the RGB.
\end{enumerate}

\begin{figure*}
\centering
\resizebox{15cm}{!}{\includegraphics{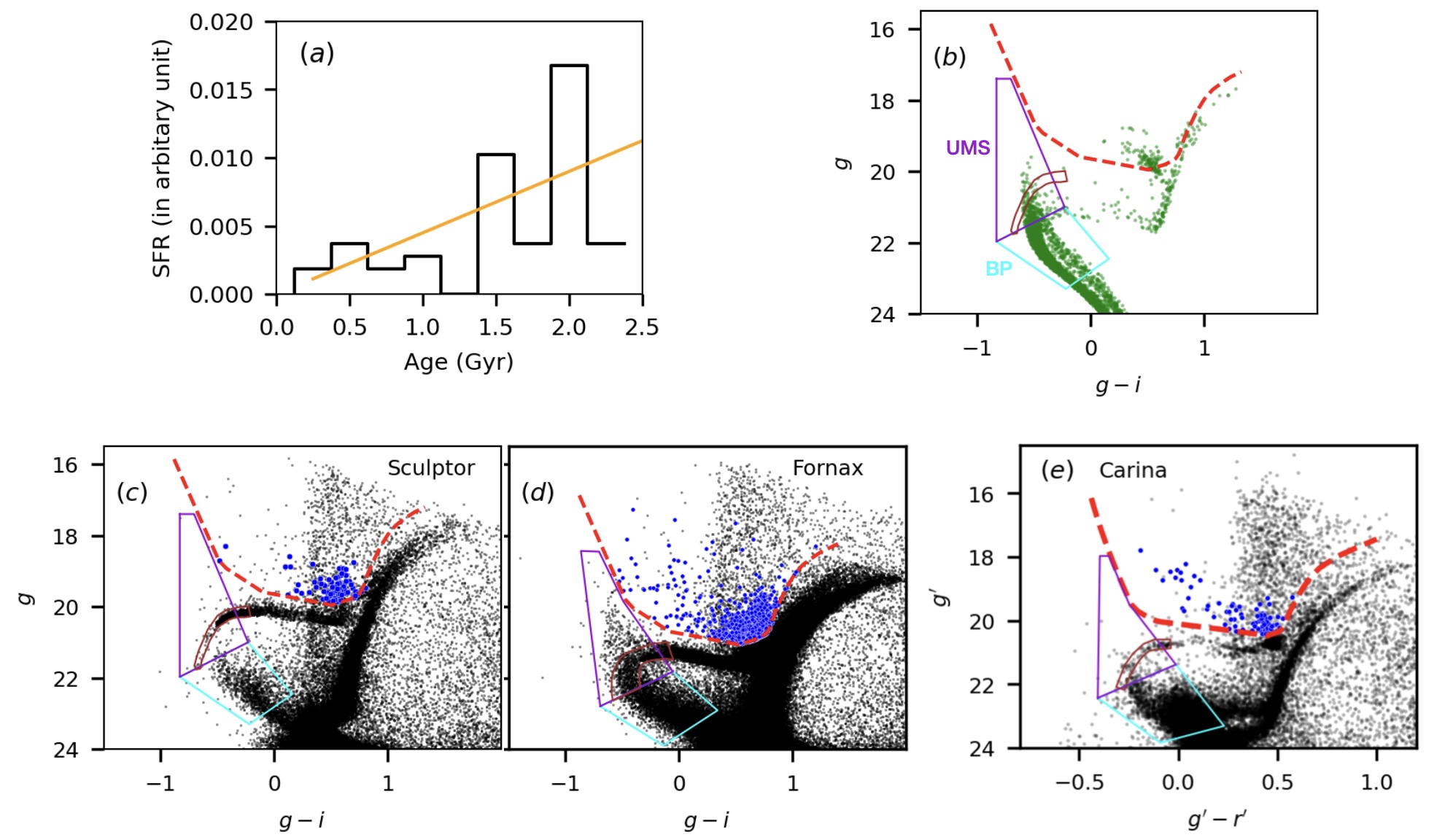}}
\caption{Stellar populations in simulations and dSphs.
Panel ($a$): Recent SFH of Sculptor predicted by simulations made in Paper~III (black lines), while the orange line is a linear fit to the SFH. 
Based on the latter fitted SFH, we simulated a synthetic CMD shown as the green points in panel~$(b)$. This synthetic CMD will be compared with Sculptor's CMD (within $r_{\rm ell}=1.1$~deg) from the Dark Energy Survey (DES) in Panels~($c$), with its CHeB candidates (blue dots) superposed.
In both panels, we define some regions for statistics: the red-dashed line for our CHeB selection condition; cyan polygons for the blue plume(BP) region;  violet polygons for the upper main sequence (UMS) region; brown polygons for HB region that to be exclude when counting inside UMS.  Panels~($d$) and ($e$) show the same content as in panel~$(c)$ but for Fornax and Carina, respectively, and the polygons are revised accordingly.
}
\label{fig:SFcheck}
\end{figure*}

\subsubsection{The upper main sequence in Fornax and Sculptor\label{sec:ums}}

When comparing the synthetic CMD (see Sect.~\ref{sec:recentSF} for details) to Sculptor's CMD (i.e., the panels ($b$) to ($c$) of Figure~\ref{fig:SFcheck}), 
we may notice there are numerous upper main sequence (UMS) stars predicted by the theory, while there are only a few in Sculptor. We established our statistics by defining a UMS selection on the CMD and comparing the counts to that of CHeB. According to the result in Table~\ref{tb:chebratios}, the observation shows a lack of UMS stars by at least factor of 3.
This could be an indication that the CHeB candidates are not young stars. As reference, we examined the situation in Fornax and Carina, which are known to have experienced star formation in the recent 2 Gyr; with particularly young stars of 150-Myr old identified in Fornax \citep{DeBoer12Fornax,Saviane2000,Stetson1998}. Regions of selection and statistical results are shown in Figure~\ref{fig:SFcheck} and Table~\ref{tb:chebratios}, respectively. Surprisingly, we found that Fornax and Carina show the same lack of UMS stars as in Sculptor. 

We have tried to vary SHF and simulate synthetic CMD (as described in Sect.~\ref{sec:recentSF}) in order to solve the discrepancy of lacking of UMS.  However, we found this is impossible because to explain the distribution of high mass CHeB stars (i.e., $>2.04~M_{\odot}$), we would need stars younger than $\approx$~700 Myr (see Figure~\ref{CHeBages} and Figure~\ref{CHeBagesA}) and the synthetic CMDs below this age always predict more UMSs than CHeBs. 

To reconcile the observations in Fornax with theory, we may envisage to
assume a non-standard initial mass-function (IMF) that is truncated at the low mass end, at about
1.8~M$_\odot$.
If there were any star formation taking place during the infall in a dSph, its gas 
may be strongly affected by the perturbation introduced by ram pressure.  
A recent work by \citet{Hennebelle2024}  suggests that star formation could be inefficient towards the low mass range in high Mach number environments, in other words, strongly perturbed gas. This theoretical prediction 
may explain 
the lack of UMS stars in Fornax, Sculptor and other dSphs. 

In this work, we have demonstrated that there are consistent stellar populations in Fornax, Carina, and Sculptor. 
Specifically, they all exhibit very similar number ratios of CHeB-to-UMS stars. Given the fact that Fornax has been forming stars very recently, the possibility of a recent star formation in Sculptor cannot be ruled out. 
Sculptor is almost ten times less massive than Fornax \citep{Hammer2019}, as also supported by the UMS counts in Table~\ref{tb:chebratios}. This may explain why the UMS stars in Sculptor cannot be recognized visually.

\subsection{CHeB candidates as indicators of a recent star formation}
\label{sec:recentSF}

If the CHeB candidates in Sculptor are genuine young stars due to a recent star formation, 
we should expect to detect the imprint of their MS counterparts in the BP to be consistent with a normal recent star formation. This could be indicated by the ratio of stellar counts between the CHeB population and the BP population. In particular, the number of CHeB stars found for Sculptor should predict more than a thousand stars in the BP.
To make this consistency check, we may compare the observations to a synthetic CMD, after assuming a recent SFH. 

In Paper III, we carried out a simulation dedicated to simulate the infall of Sculptor that also predicts the star formation induced by ram pressure during a recent infall. Figure~\ref{fig:SFcheck}$a$ shows the corresponding predicted SFH from paper III (see the black histogram). The SFH can be approximated by a linear and declining star formation activity in the last 2.5 Gyr (i.e., the orange line in Figure~\ref{fig:SFcheck} $a$) hereafter denoted as the "predicted" SFH.

We used \texttt{BaSTI} to simulate a synthetic CMD of a recent star formation by assuming the predicted SFH between 0.5 to 2.5 Gyr, a Kroupa IMF \citep{Kroupa2001}, and an averaged [Fe/H]~$=-1.5$. The simulated synthetic CMD is shown in Figure~\ref{fig:SFcheck} $b$. For improved visualization, we re-scaled the synthetic CMD to have CHeB counts of 93 (i.e., all points above red-dashed line in the figure), which is similar to that of Sculptor from {\it Gaia} dataset.

To make this comparison possible, we had to extend our dataset for the faint MS stars by using the deep photometry from the Dark Energy Survey (DES). 
We obtained stars within 1.1 degree elliptical radius centered on Sculptor from the DES data release 2 (DR2). The star-galaxy separation was done using the parameter \texttt{extended\_class\_coadd} = [0,1] for stellar sources, following the official document. 
We may define a BP region using a polygon (in cyan color) on the CMD (Figure~\ref{fig:SFcheck}), after avoiding the possible contamination from HB stars in the bright end and possible confusion due to photometric errors in the faint end.

Then, we go on to count the number of CHeB and BP stars, in both observed and  synthetic CMD, and the values are listed in Table~\ref{tb:chebratios}.  
As we have demonstrated, the {\it Gaia} detection of CHeB stars has few contamination, so we may directly compare it to the CHeB counts from the simulated CMD. For the BP population observed by DES, the MW contamination is likely very small after examining the CMD; thus, we did not apply any correction to these BP counts (listed in Table~\ref{tb:chebratios}).
The result shows that the number ratios are similar, indicating that the CHeB population detected in {\it Gaia} data is compatible with recent star formation within the last 2.5 Gyr after assuming the predicted SFH from simulation of a recent infall of Sculptor. Table~\ref{tb:chebratios} shows that the ratio obtained from observations is slightly larger than that from the simulated CMD. Such a difference could be due to the input IMF or possibly related to the incompleteness in the observation of DES in the magnitude range of our BP selection. The latter is especially relevant when considering that our statistics from observations also include the crowded stellar field in the center of Sculptor.

We  extended this statistical test to the other dSphs, namely, Sextans, UMi, Draco, Carina, and Fornax, with the results listed in Table~\ref{tb:chebratios}. We adopted the deep photometry of these three objects from the observations by \citet{Munoz2018}. The latter were carried out with MegaCam system, which is a different photometric system from DES; thus, we have carefully defined a BP region, which is equivalent to the BP definition in Figure~\ref{fig:SFcheck}, on the CMD in MegaCam filter system (see Appendix~\ref{sec:otherdSphs} for the definition of these photometric systems).

\renewcommand{\arraystretch}{1.2}
\begin{table} 
\caption{Statistics of CHeB, BP, and UMS populations.}
{\centering
\begin{tabular}{lccccc}
\hline \hline
     & $N_{\rm CHeB}$ & $N_{\rm BP}$ & $\frac{N_{\rm CHeB}}{N_{\rm BP}}$  & $N_{\rm UMS}$ & $\frac{N_{\rm CheB}}{N_{\rm UMS}}$ \\ \hline 
Synthetic CMD & 116 & 2010 & 0.057 & 240 & 0.48  \\ 
Sculptor & 90 & 1169  & 0.077 & 55  &  1.64   \\ 
Sextans & 38 & 543 & 0.069 & -- & -- \\
UMi & 54 & 795 & 0.067  & -- & --   \\
Draco & 36 & 589 & 0.061 & -- & -- \\ 
Carina &  95 & 8065  & 0.012 & 77  &  1.23 \\
Fornax &  964 & 4794  & 0.201 & 516  &  1.87  \\  
\hline
\end{tabular}}\\
Notes: $N_{\rm CHeB}$ is equal to $N_{\rm net}$ that is quoted in Table~\ref{tab:CHeBstat} for dSphs.
\label{tb:chebratios}
\end{table}
\renewcommand{\arraystretch}{1.}

\subsection{Recent star formation in dSphs  during their recent infall into the MW halo}
Figure~\ref{CHeBages} 
indicates a recent star formation from $\sim 0.4$ to 2 Gyr after accounting for the young stars detected in Sculptor. 
This is almost consistent with a single star formation burst 
that occurred about 1 Gyr ago.  
A short starburst does not 
necessarily result in a narrow range of metallicity, as demonstrated 
by the case of Boo I, for which a single starburst with a duration of only 50 Myr produced
a spread in metallicities of about 2\,dex \citep[see][Figures 3 and 4]{martina}. 
In spite of the large errors associated 
to our chemical abundance measurements in Sculptor, 
the spread in metallicity seems to be real. Although we cannot formally
exclude the possibility that the stars share  
a single metallicity ([Fe/H]=$-1.2\pm 0.33$), 
this would not explain the [Mg/Fe] and [Ca/Fe] trends shown in Figure~\ref{mgfe}. 
The typical timescale for SN Ia enrichment depends on the star formation rate and on the IMF. \citet{MR01} estimated that for a single starburst, this can be as short a timescale as 40--50 Myr. 
The fact that the young population displays a trend in  [Mg/Fe] and [Ca/Fe] implies that
this timescale has been very short indeed.
\citet{Zhu2024} conducted dedicated simulation to investigate the SF processes in gas-rich dwarf galaxies under conditions of ram pressure striping (RPS). Their results suggest that bursty SF appears to be a typical feature during all their simulation time until gas is completely removed from galaxies. Such bursty SF has also been observed in Fornax by \citet{Rusakov2021}, and seems to be consistent with the SF induced by RPS \citep[e.g., see the discussion in][]{Yang2022}.

If the spread in Figure~\ref{mgfe} were due to measurement errors alone, we 
 would not expect to see any trend among the abundance ratios. 
This is especially true for  [Fe/H] and [Ca/Fe], for which a Kendall's $\tau$ statistics gives a 
99.999\% probability that they are anti-correlated. It is difficult to dismiss these findings, and they need 
to be explained if they result from a recent star formation hypothesis.
CHeB candidates of different metallicity do not show a 
defined age-metallicity relation (see the {\bf top} panels of Figure~\ref{CHeBages}). 
The ages and metallicities for the CHeB candidates in Sculptor, 
whether we take [Fe/H] or [Ca/H], 
are not correlated; however, they are expected to be undergoing a star formation event. 
The past history of the dSph hosts of CHeB candidate stars, as described in 
Papers I, II, and III, provides
a plausible interpretation of the observations.

MW dSph progenitors are gas-rich galaxies, namely, dwarf irregular galaxies (dIrrs) that have  recently ($<$ 3 Gyr ago) been accreted by the MW (see Paper I). During their infall, they lost their gas due to the ram pressure of the MW halo corona. This fully transformed them from gas-rich dwarfs at equilibrium to gas-free dwarfs sufficiently out of equilibrium to explain their large velocity dispersions (see Papers II and III). The simulations carried out in Paper III  showed that before being removed, the gas is affected by considerable turbulence through shocks and compression\footnote{A video realized during the Paper III modeling of a recent infall of the Sculptor dSph progenitor is available at: https://www.youtube.com/watch?v=SwxSdmfQis4}. The latter phenomenon led to star formation, while the former ensures that the whole gas medium has been literally shaken, providing star formation within a mixture of gas-rich clouds with various metallicities. This is consistent with the absence of  a defined age-metallicity relation in Figure~\ref{CHeBages}, as well as with the observed metallicity spread. 
This may also explain the different metallicity distributions between the old population
(age $> 6$\,Gyr) and the young population (age $< 2.5$\,Gyr), the latter lacking the most
metal-poor tail and extending to slightly higher metallicities. The fact that there
is a wide metallicity overlap between the two populations implies that the
gas that fuelled the starburst  was not chemically homogeneous.

The role of ram pressure is twofold: i) removing the gas in infalling dIrrs and ii) mixing and compressing gas from outer region (metal poor) towards inter region (metal rich) in the galaxy. If the gas is compressed to a high enough  density, there will be star formation processes ignited.  
Our finding that part of the CHeB candidates in dSphs are possibly young stars is consistent with the orbital history of dSphs that are revealed by {\it Gaia} proper motions. 
Our results support a very recent star formation associated to the recent infall of dSph progenitors (see Figure~\ref{fig:SFcheck}-a). Simulations from Paper III have successfully reproduced the observed properties for Sculptor assuming a recent infall. They predicted a very small fraction ($\ll$~1\% of the total stellar mass) of stars formed in the last 1.5 Gyr during the infall of the simulated dwarf galaxy. 
Taking the estimated mass of the CHeB candidates, we may infer the total mass formed in Sculptor after assuming an initial mass function (IMF). Given the 91 CHeB candidates with masses ranging from 1.8 to 3.0~$\msun$, 
by inverting the Kroupa IMF \citep{Kroupa2001}, for example,  we can derive the corresponding stellar mass formed during the same time interval; this is 7200~M$_\odot$, at a star formation rate (SFR) of $3.6\times 10^{-6}$~M$_\odot$/yr over 2.5 Gyr.
If we adopt a total stellar mass of $5.1\times 10^6$~M$_\odot$ for Sculptor (see Paper II), the mass fraction of young stars is 0.15\%. \\

\section{Conclusions}
In  this work, we have identified a  population of apparently young, intermediate mass CHeB stars, which has been distinguished from MW foreground stars using {\it Gaia} DR3 PMs. They are found in several dSphs,  including Sculptor, UMi, and Sextans, which were previously 
thought to possess only old stars, as well as Draco. For the latter, the existence of a young population
was claimed by \citet{Aparicio2001}, but confuted by \citet{Mapelli2007}.  
Their ages and masses have been computed using a grid of evolutionary tracks and they 
have been confirmed using isochrones,  for  a sub-sample with spectroscopic metallicities. \\

Two possible channels have been considered to explain the observations of CHeB candidates. It could be that they are evolved BSSs or, alternatively, they could have been formed during the star formation induced by ram pressure at a recent epoch; namely, when gas-rich dSph progenitors entered in the MW halo and its corona.\\

The former hypothesis faces serious difficulties, because the stars had to have gone through very rare events 
of coalescence of at least three stars to form a single BSS. This conclusion is very robust since many CHeB candidates have masses in excess of  1.8\,$\msun$, while dSph turnoff mass stars have to be smaller then 0.9 $\rm M_\odot$. This is because 95.5\% of Sculptor stars are older than 8 Gyr, leading to masses smaller than 0.9 M$_{\odot}$, which only allows BSS masses of less than 1.8 M$_{\odot}$ (much smaller than what we have observed). Furthermore, the BSS hypothesis is not sufficient to explain why some CHeB candidates have larger metallicities than RGB stars, as well as why none of them share the lowest metallicities of RGB stars.\\

All the CHeB star candidates are then considerably younger than the sample of \citet{hill19} ($< 1$\,Gyr compared to 6\,Gyr, the youngest ages in the RGB sample). Such recent star formation events are consistent with a recent infall of dSph progenitors that are expected to be gas-rich dIrrs (see paper I). During their infall, the gas was very turbulent and its whole content permanently shaken before being removed by the ram pressure exerted by the Galactic corona (see Paper III). Such a hypothesis is also supported by the comparison of number counts of CHeB stars with those UMS stars in Sculptor and Fornax. 
The existence of a young stellar population in Fornax has been robustly established.
This peculiar series of processes explains the large velocity dispersions of dSphs (see Papers II and III). 
This approach is also consistent with the chemical analysis presented here; specifically, the absence of an age-metallicity relation and the spread in metallicity despite of the short-duration of these star formation events.
If confirmed by further observations, this population of young-and-intermediate mass 
stars may 
be the last witnesses of very recent star formation events in dSphs, 
including in those reputed to possess only a very old star population.

\begin{acknowledgements}
We thank Dr. Santi Cassisi for the insightful discussions, his help in using the \texttt{BaSTI} and simulating the synthetic CMD.
We thank Eva Grebel for her explanations of many former CMD studies of Sculptor.
We are grateful to Dr.\ Yang Chen for his help on the use of the PARSEC stellar evolutionary tracks and the YBC database of stellar bolometric corrections. 
We thank Dr.\ Monique Spite, Dr.\ Marcel S.\ Pawlowski, Dr.\ Haifeng Wang, Mr.\ Yongjun Jiao and Dr.\ Hefan Li for their helpful discussions. 
We are also grateful to Prof. E. Tolstoy and Prof. A. Sk{\'u}lad{\'o}ttir for their very useful comments on our paper. 
We are also grateful to Prof. A. Sills for providing us with the evolutionary tracks of BSS.
We are grateful for the support of the International Research Program Tianguan, which is an agreement between the CNRS in France, NAOC, IHEP, and the Yunnan Univ. in China.
This work presents results from the European Space Agency (ESA) space mission {\it Gaia}. {\it Gaia} data are being processed by 
the {\it Gaia} Data Processing and Analysis Consortium (DPAC). Funding for the DPAC is provided by national institutions, 
in particular the institutions participating in the {\it Gaia} MultiLateral Agreement (MLA). 
The {\it Gaia} mission website is https://www.cosmos.esa.int/gaia.
PB and EC gratefully acnkowledge 
support from the European Research Council
through Advanced Grant  835087 -- ``SPIAKID''
and from the French National Research Agency (ANR) through project 
``Pristine'' (ANR-18- CE31-0017). 

This project used public archival data from the Dark Energy Survey (DES). Funding for the DES Projects has been provided by the U.S. Department of Energy, the U.S. National Science Foundation, the Ministry of Science and Education of Spain, the Science and Technology FacilitiesCouncil of the United Kingdom, the Higher Education Funding Council for England, the National Center for Supercomputing Applications at the University of Illinois at Urbana-Champaign, the Kavli Institute of Cosmological Physics at the University of Chicago, the Center for Cosmology and Astro-Particle Physics at the Ohio State University, the Mitchell Institute for Fundamental Physics and Astronomy at Texas A\&M University, Financiadora de Estudos e Projetos, Funda{\c c}{\~a}o Carlos Chagas Filho de Amparo {\`a} Pesquisa do Estado do Rio de Janeiro, Conselho Nacional de Desenvolvimento Cient{\'i}fico e Tecnol{\'o}gico and the Minist{\'e}rio da Ci{\^e}ncia, Tecnologia e Inova{\c c}{\~a}o, the Deutsche Forschungsgemeinschaft, and the Collaborating Institutions in the Dark Energy Survey.

The Collaborating Institutions are Argonne National Laboratory, the University of California at Santa Cruz, the University of Cambridge, Centro de Investigaciones Energ{\'e}ticas, Medioambientales y Tecnol{\'o}gicas-Madrid, the University of Chicago, University College London, the DES-Brazil Consortium, the University of Edinburgh, the Eidgen{\"o}ssische Technische Hochschule (ETH) Z{\"u}rich,  Fermi National Accelerator Laboratory, the University of Illinois at Urbana-Champaign, the Institut de Ci{\`e}ncies de l'Espai (IEEC/CSIC), the Institut de F{\'i}sica d'Altes Energies, Lawrence Berkeley National Laboratory, the Ludwig-Maximilians Universit{\"a}t M{\"u}nchen and the associated Excellence Cluster Universe, the University of Michigan, the National Optical Astronomy Observatory, the University of Nottingham, The Ohio State University, the OzDES Membership Consortium, the University of Pennsylvania, the University of Portsmouth, SLAC National Accelerator Laboratory, Stanford University, the University of Sussex, and Texas A\&M University.

Based in part on observations at Cerro Tololo Inter-American Observatory, National Optical Astronomy Observatory, which is operated by the Association of Universities for Research in Astronomy (AURA) under a cooperative agreement with the National Science Foundation.
\end{acknowledgements}

\section*{Data Availability}
The catalogs of the PM-selected and CheB candidates are available upon request.



\bibliographystyle{mnras}
\bibliography{biblio} 





\begin{appendix}

\section{Detection of CHeBs in other dSphs}
\label{sec:otherdSphs}
\begin{figure*}
{
\centering
\resizebox{14cm}{!}{\includegraphics{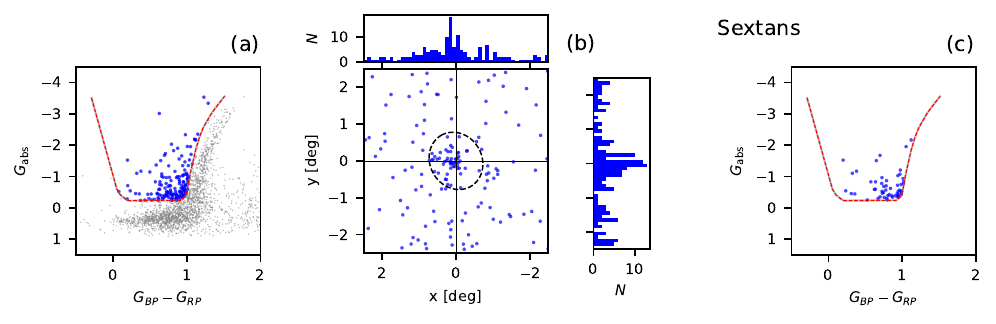}}\\
\resizebox{14cm}{!}{\includegraphics{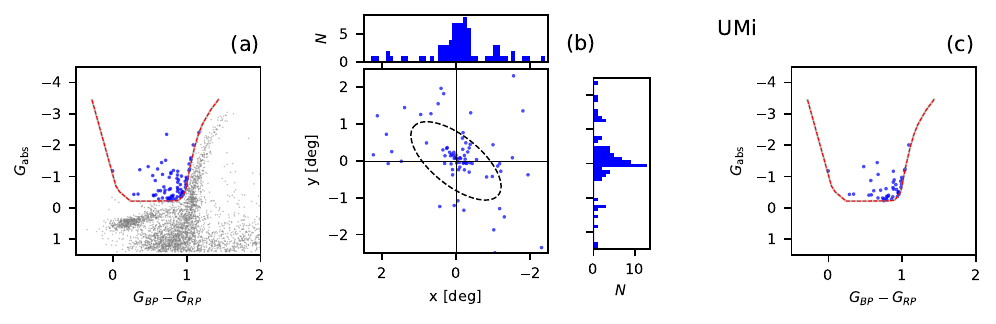}}\\
\resizebox{14cm}{!}{\includegraphics{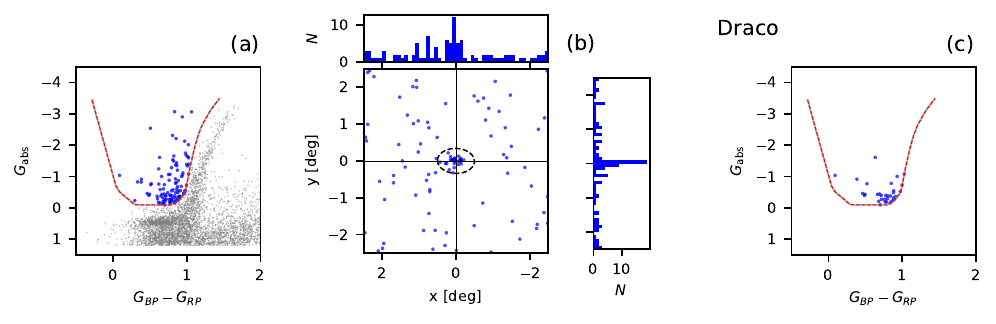}} \\
\resizebox{14cm}{!}{\includegraphics{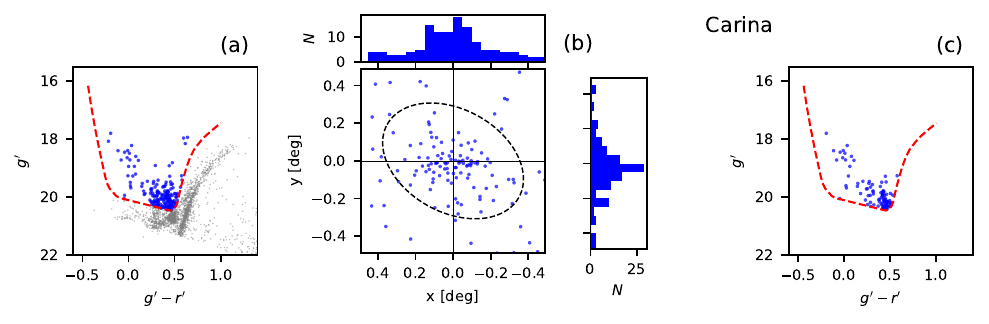}} \\
\resizebox{14cm}{!}{\includegraphics{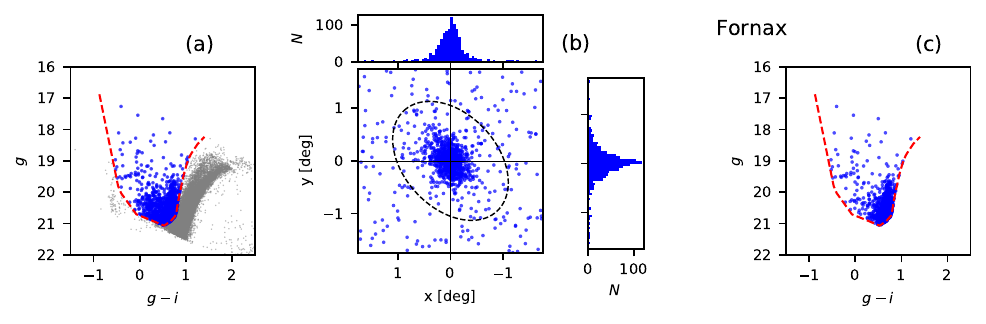}} \\
}
\caption{Selection of CHeB candidates and their spatial distribution for each dSphs. Panels and symbols are the same as in Figure~\ref{fig:sculstat}.}
\label{theCHeB}
\end{figure*}

\begin{figure}
\centering
\resizebox{8cm}{!}{\includegraphics{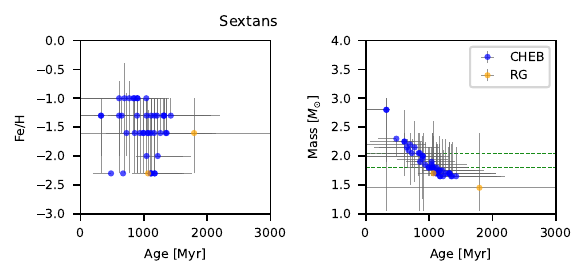}} \\
\resizebox{8cm}{!}{\includegraphics{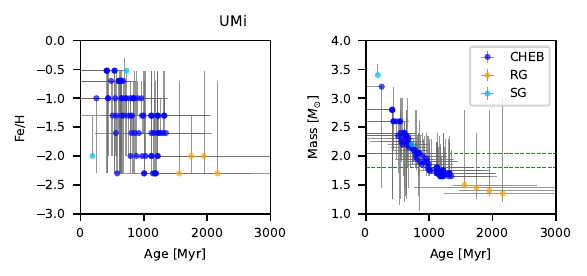}} \\ 
\resizebox{8cm}{!}{\includegraphics{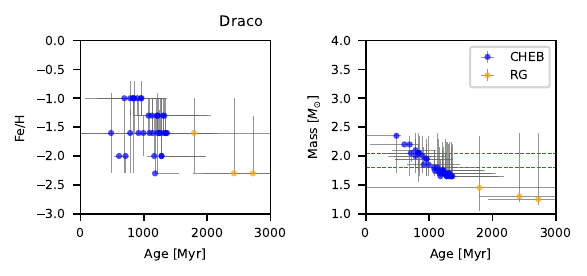}} \\
\resizebox{8cm}{!}{\includegraphics{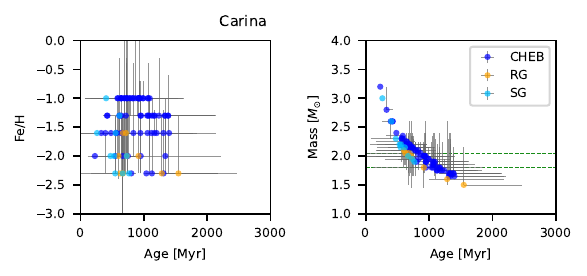}} \\ 
\resizebox{8cm}{!}{\includegraphics{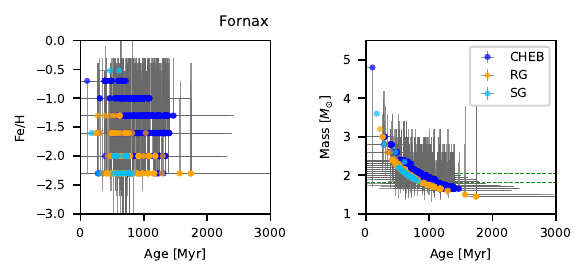}} \\
\caption{Age-metallicity ({\it Left}) and age-mass ({\it Right}) relations for CHeB stars in their different evolutionary phases (indicated by color explained in the legend) for each dSph, respectively. The two dashed lines in the right panel indicate the masses of 1.8, 2.04~$M_{\odot}$, respectively. 
}
\label{CHeBagesA}
\end{figure}

 For Draco, UMi, and Sextans dSphs, the methodology to derive CHeB counts is very similar to that described for Sculptor (see Sect.~\ref{sec:chebdetection}). The precise locations of the 2-Gyr limit of RGB are different for each dSphs because we computed the corresponding evolutionary tracks according to their Galactic extinctions, respectively. 
The detection of CHeBs in each dSphs are described by Figure~\ref{theCHeB}, and their stellar parameters are shown in Figure~\ref{CHeBagesA}.
A special remark for UMI: following the algorithm of defining the $r_{
\rm ell}$ in Sect.~\ref{sect:step2}, we found a relative large value of $r_{\rm ell} = 2.5$~deg, which is possibly related to the extended stellar halo by \citet{Sestito2023}. Nonetheless, we choose to fix $r_{\rm ell}=1.48$~deg, i.e., 5$r_{\rm half}$, to deduce the statistics in Table~\ref{tab:CHeBstat}, in order to keep consistency with the analysis of other dwarf galaxies.
\\

 However, {\it Gaia} data are insufficiently deep to properly detect the CHeB population in Fornax and Carina.
Fornax is located at 141 kpc, i.e., a distant module $m-M_0 = 20.72$, i.e., one magnitude fainter than Sculptor. 
Thus, when applying our selection of CHeB candidate, the faint limit ($G=20.65$) of our CHeB condition in Fornax is too close to the limiting magnitude of {\it Gaia}, $G=20.8$ as we have adopted. 
Then, "Gaia" CHeB candidates in Fornax are likely contaminated by  RGB or AGB stars because the photometric errors in color are likely too large. The only way to solve this problem is to replace the {\it Gaia} photometric data by more accurate data. 
DES is a good choice because it provides not only a high quality photometry but also a wide coverage on sky.

We retrieved a catalog of sources from the DES data release 2 (DR2) within 2.5-degree radius centered on Fornax, using \texttt{|extended\_class\_coadd|<=1}, as recommended by the DES-DR2 official document\footnote{\hyperlink{https://des.ncsa.illinois.edu/releases/dr2/dr2-faq}{https://des.ncsa.illinois.edu/releases/dr2/dr2-faq}}. 
To combine the DES and {\it Gaia} data, we cross-matched the {\it Gaia} PM-selected sample of Fornax to the DES star catalog, and obtained a "GaiaPM-DES" catalog, which is based on the DES photometry, and for which the MW huge contamination has been filtered by the {\it Gaia} PMs. 

To select CHeB candidate, we have transformed the CHeB selection condition that is defined in {\it Gaia} photometric system into the DES photometric system, using the photometric relationships provided by {\it Gaia} EDR3 (Sect. 5.5.1 in ``{\it Gaia} early data release 3 documentation"): 
\begin{align}
        G - g &= 0.2199 -0.6365\,x -0.1548\,x^2 + 0.0064\,x^3 \ ,
 \label{eq:Gtog} \\
    G - i &= -0.293 + 0.6404\,x -0.09609\,x^2 \nonumber \\
    & \quad\quad -0.002104\,x^3\ , \label{eq:Gtoi} \\
    G - r &= -0.09837 + 0.08592\,x -0.1907\,x^2 -0.1701\,x^3 \nonumber \\
    & \quad\quad + 0.02263\,x^4 \ , \label{eq:Gtor}
\end{align} 
where $x = G_{BP} - G_{RP}$. 
The CHeB selection condition in DES photometric system is shown by the red-dashed line in Figure~\ref{theCHeB}, where the Fornax' CHeB candidates are indicated by blue dots. \\

For Carina, we have similar difficulties as in Fornax based on only the {\it Gaia} data, due to its relative large distance of 106 kpc ($m-M_0 = 20.13$). Because Carina is not covered by DES, we have to adopt the observation from the MegaCam survey by \citet{Munoz2018}, as an accurate photometric alternative. The MegaCam observation on Carina covers $0\fdg7\times0\fdg7$, and was carried out using the MegaCam at the Canada France Hawaii Telescope (CFHT), which adopted a slightly different filter system than DES. In order to apply the same CHeB selection to MegaCam filters, we have translated the selection line from Gaia's photometric system into that of DES (using Eqs.~\ref{eq:Gtog} and \ref{eq:Gtor}), then into MegaCam photometric systems, i.e., $g^\prime$ and $r^\prime$ by applying the other two formula below\footnote{\url{https://www.cadc-ccda.hia-iha.nrc-cnrc.gc.ca/en/megapipe/docs/filtold.html}}: 
{\small
\begin{align}
        g^\prime &= g - 0.153\,(g-r) ,\label{eq:gtogp} \\
    r^\prime &= r - 0.024\,(g-r). \label{eq:rtorp}
\end{align} 
}
Then, we cross-matched the MegaCam catalog with Carina's {\it Gaia} PM-selected sample to obtain a "GaiaPM-MegaCam" catalog, and finally obtained the GaiaPM-MegaCam CHeB candidates, as shown in Figure~\ref{theCHeB}.

After obtaining the CHeB Candidates of Carina and Fornax, we made the statistics on the candidates using the same methods described in Sect.~\ref{sec:CHeBselection}, and the results are listed in Table~\ref{tab:CHeBstat}. Note that, to derive the stellar parameters of CHeB candidates, we have generated the corresponding reddened PARSEC evolutionary tracks in DESCam and MegaCam systems that are dedicated to Fornax and Carina, respectively.

\FloatBarrier

\section{Solutions for stellar parameters }
\label{sec:agealgo}

An evolutionary track of a star is initialized by a stellar mass and a metallicity (which is, of cause, simplified for our main interests), and is made of a sequence of points in time (i.e., age) that describe the corresponding parameters of the evolutionary track of the star.
When we follow the latter in the CMD, it evolves at different speeds. 
For the $i$-th point of the track at time $t^i$, we may calculate an evolutionary speed in both magnitude and color, as
{\small
\begin{align}
v^i_{\rm mag} &= {|m^{i+1} - m^{i-1}| \over  t^{i+1} - t^{i-1}}\ ,\\
v^i_{\rm col} &= {|c^{i+1} - c^{i-1}| \over t^{i+1} - t^{i-1}} \ ,
\end{align}}
where $m= G_{\rm abs}$ and $c=G_{BP}-G_{RP}$ in the {\it Gaia} photometric system.
Assuming a track that is broadened by star photometric uncertainties  ($m\pm\Delta m$ and $c\pm\Delta c$), we may calculate the time that the track needs to pass through twice of the error bars, respectively, as
{\small
\begin{align}
\tau^i_{\rm mag} &= 2\, \Delta m / v^i_{\rm mag},\\
\tau^i_{\rm col} &= 2\, \Delta c / v^i_{\rm c},
\end{align}}
Here, we adopted 1-sigma errors for $\Delta m$ and $\Delta c$ in magnitude and color, respectively, when working with {\it Gaia} photometry, while a "3 sigma error" was adopted when working on the DES and MegaCam dataset. The latter two datasets provide very accurate photometry, namely, small errors, while the grid of the PARSEC evolutionary track currently adopted is too large when compared to the small error bars obtained from both deep observations. Thus, we have to enlarge the searching window when looking for possible solutions in evolutionary tracks.
In practice, the photometric errors for a given star could cover not only one or more points of the same track, but also cover one or more different evolutionary tracks. 
Our goal is to retrieve the most probable evolutionary track.
Thus, we need to average all points of the same track that are covered by the photometric errors, 
by evaluating the mean $\tau_{\rm mag}$ and $\tau_{\rm col}$ of the points, together with the mean of the stellar parameters (see Table~\ref{agealgo} for example). 
The column $N_{\rm shot}$ tells the number of points of a track that is covered by the photometric errors.
Taking the multiplication of $\tau_{\rm mag}\tau_{\rm col}$ as the final weighting, 
we sort tracks that are covered by $\Delta m$ and $\Delta c$, and identify the most probable one with the maximal \emph{elapsed-time}.  

As the final solution for a given star, we report its most probable evolutionary track, together with the range of tracks that are indexed by stellar mass. In another words, we report the parameters of the tracks of minimal or maximal stellar mass. In doing this, we keep the physical consistency in between parameters. 
In the PARSEC library, evolution phases are normalized into a number $p$, ranging from 1 to 15, and each integer point corresponds to the starting of a new phase of the evolution. Our dictionary is $1<p({\rm PMS})<5<p({\rm MS})<6<p({\rm SG})<8<p({\rm RGB})<11<p({\rm CHeB})<14<p({\rm AGB})<15$.

\begin{table*}
\caption{Examples of the estimation of stellar parameters}
\label{agealgo}
\small
\begin{tabular}{lccccrrrrrr}
\multicolumn{5}{l}{{\bf Example 1}: A star in Sculptor: $G_{\rm abs}=-0.7984 \pm 0.0013 $, $G_{BP}-G_{RP}=0.919 \pm 0.024$} \\
\multicolumn{5}{l}{Possible solutions:} \\ \hline
Track&Mass & [Fe/H] & Age & Phase && $N_{\rm shot}$& $\tau_{\rm mag}$ & $\tau_{\rm col}$ & $\tau_{\rm mag}\tau_{\rm col}$  \\
 &$M_{\odot}$ &  & (Myr) &   & & & (Myr) & (Myr) & (Myr$^2$) \\\hline
1& 2.25 & $-1.602$ & 543.5 &CHeB && 3 & 14.2 & 7.0 & 99.3 \\
2& 2.10 & $-1.000$ & 830.5 &CHeB && 1 & 0.289 & 6.5 & 1.87 \\
3& 2.15 & $-2.000$ & 572.6 &CHeB && 1 & 0.0858 & 9.0 & 0.776 \\
4& 2.05 & $-2.000$ & 644.3 &CHeB && 1 & 0.0632 & 11.0 & 0.717 \\
5& 2.35 & $-1.301$ & 511.8 &CHeB && 1 & 0.127 & 4.5 & 0.565 \\
6& 2.00 & $-2.301$ & 670.1 &RG && 1 & 0.0103 & 2.8 & 0.0286 \\
7& 2.00 & $-2.000$ & 687.5 &CHeB && 1 & 0.00835 & 1.9 & 0.016 \\
8& 1.80 & $-1.602$ & 1111.8 &CHeB && 3 & 0.0297 & 0.39 & 0.0117 \\
9& 2.10 & $-2.301$ & 587.3 &RG && 1 & 0.00774 & 1.5 & 0.0113 \\
10& 1.75 & $-1.602$ & 1206.0 &AG && 1 & 0.00221 & 0.23 & 0.000514 \\ 
\hline \\
\multicolumn{5}{l}{{\bf Example 2}:\ Astar in Sculptor: $G_{\rm abs}=-0.5177 \pm 0.0016 $, $G_{BP}-G_{RP}=0.942 \pm 0.027$} \\
 \multicolumn{5}{l}{Possible solutions:} \\  \hline
Track & Mass & [Fe/H] & Age & Phase && $N_{\rm shot}$& $\tau_{\rm mag}$ & $\tau_{\rm col}$ & $\tau_{\rm mag}\tau_{\rm col}$  \\
& $M_{\odot}$ &  & (Myr) &   &&& (Myr) & (Myr) & (Myr$^2$) \\\hline
1& 1.40 & $-2.301$ & 1916.1 &RG && 1 & 0.0641 & 14 & 0.882 \\
2& 1.55 & $-2.000$ & 1415.4 &RG && 1 & 0.0499 & 11 & 0.573 \\
3& 1.95 & $-2.000$ & 738.7 &CHeB && 1 & 0.0578 & 9.6 & 0.558 \\
4& 1.55 & $-2.301$ & 1399.5 &RG && 1 & 0.0499 & 11 & 0.549 \\
5& 1.60 & $-2.301$ & 1271.5 &RG && 1 & 0.0454 & 9.9 & 0.448 \\
6& 2.35 & $-1.000$ & 536.1 &CHeB && 1 & 0.0835 & 4.5 & 0.374 \\
7& 1.65 & $-2.000$ & 1176.5 &RG && 1 & 0.04 & 7.8 & 0.31 \\
8& 2.05 & $-1.301$ & 709.8 &CHeB && 1 & 0.0404 & 5.9 & 0.237 \\
9& 1.50 & $-2.000$ & 1563.7 &RG && 1 & 0.0416 & 4.3 & 0.177 \\
10&1.85 & $-2.000$ & 847.6 &RG && 1 & 0.0172 & 4.2 & 0.0719 \\
11&1.85 & $-1.602$ & 883.0 &RG && 1 & 0.0152 & 2.9 & 0.0437 \\
12&1.75 & $-1.301$ & 1272.8 &CHeB && 1 & 0.0263 & 1.1 & 0.0285 \\
13&1.95 & $-1.602$ & 765.3 &RG && 1 & 0.0116 & 2.1 & 0.0246 \\
14&1.70 & $-1.301$ & 1386.2 &CHeB && 1 & 0.0134 & 0.81 & 0.0109 \\
15&2.10 & $-1.301$ & 658.1 &RG && 1 & 0.00837 & 1.2 & 0.0102 \\
16&2.10 & $-1.602$ & 629.3 &RG && 1 & 0.00813 & 1.1 & 0.0093 \\
17&2.15 & $-1.602$ & 592.0 &RG && 1 & 0.0067 & 0.99 & 0.00667 \\
18&2.20 & $-1.301$ & 583.2 &RG && 1 & 0.00672 & 0.93 & 0.00623 \\ 
   \hline \\
\multicolumn{5}{l}{{\bf Example 3}:\ A star in UMi: $G_{\rm abs}=-2.3467 \pm 0.0009 $, $G_{BP}-G_{RP}=0.728 \pm 0.010$} \\
\multicolumn{5}{l}{Possible solutions:} \\ \hline
Track & Mass & [Fe/H] & Age & Phase && $N_{\rm shot}$& $\tau_{\rm mag}$ & $\tau_{\rm col}$ & $\tau_{\rm mag}\tau_{\rm col}$  \\
& $M_{\odot}$ &  & (Myr) &   &&& (Myr) & (Myr) & (Myr$^2$) \\\hline
1& 3.40 & $-2.000$ & 190.7 &SG && 1 & 0.000582 & 0.0025 & 1.45e-06 \\
2& 3.60 & $-1.301$ & 179.3 &SG && 1 & 0.000468 & 0.0023 & 1.09e-06 \\ 
     \hline \\
\multicolumn{5}{l}{{\bf Example 4}:\ A star in Draco: $G_{\rm abs}=-0.3985 \pm 0.0127 $, $G_{BP}-G_{RP}=0.487 \pm 0.067$} \\
\multicolumn{5}{l}{Possible solutions:} \\ \hline
Track & Mass & [Fe/H] & Age & Phase && $N_{\rm shot}$& $\tau_{\rm mag}$ & $\tau_{\rm col}$ & $\tau_{\rm mag}\tau_{\rm col}$  \\
& $M_{\odot}$ &  & (Myr) &   &&& (Myr) & (Myr) & (Myr$^2$) \\\hline
1& 1.70 & $-2.000$ & 1164.5 &CHeB && 6 & 4.2 & 51 & 214 \\
2& 1.65 & $-2.000$ & 1270.8 &CHeB && 13 & 4.37 & 47 & 203 \\
3& 1.80 & $-2.301$ & 941.1 &CHeB && 4 & 2.87 & 6.3 & 18.1 \\
4& 1.85 & $-2.000$ & 885.8 &CHeB && 2 & 1.91 & 4.3 & 8.16 \\
5& 2.00 & $-2.301$ & 666.9 &SG && 8 & 0.286 & 0.24 & 0.0694 \\
6& 2.15 & $-1.301$ & 617.3 &SG && 1 & 0.136 & 0.21 & 0.0287 \\ 
  \hline \\
\end{tabular} 
\end{table*}

\FloatBarrier

\section{BSS formed from the fusion of a triple system}
\label{triples}
The CHeB stars can be used to determine
star formation histories \citep[see e.g.,][]{Dohm-Palmer1997},
having the advantage to be brighter than MS stars.
The presence of a CHeB vertical sequence is usually
taken as evidence of the presence of a young population.
However, as pointed out by \citet{Momany2007}, this can
be ambiguous, and they quote as example the GC M\,80,
for which \citet{Ferraro1999} detected a population
of CHeB stars. Since a recent star formation
in an old GC can be ruled out, these stars have to be interpreted as
evolved BSSs. The BSSs population is very
developed in M\,80 and can explain the presence of evolved BSSs.
We point out that a crucial quantity is the masses of the CHeB stars.
The brightest CHeB stars in M\,80 are brighter than the HB, however
their mass is of the order of 1.8\,$M_\odot$. Such a mass is compatible
with the merging of two 0.9\,$M_\odot$ stars.
The fainter stars in the CHeB sequence of M\,80 have lower masses.
While for masses lower than 1.8\,$M_\odot$ a CHeB could
be either a young star or a BSSs, for larger masses the BSS
explanation requires the merging of at least three stars.

The formation of a BSS by the fusion of three stars
requires the evolution of a triple system through a common envelope phase \citep{meyer80}. Measurements of BSS masses are rare in the literature \citep[see e.g.][and references therein]{raso19}.
In the globular cluster 47\,Tuc (= NGC 104, metallicity around $-0.7$), \citet{raso19} could measure masses for 23 BSSs,
of these, five have masses that are formally larger than twice the turnoff mass of the cluster.
However the errors are large enough that they are all consistent with being below this limit.
If all the five stars were truly with masses above twice the turnoff mass, this would give
a fraction of 22\% of BSSs descendants from triple systems. Inspection of figure 8 of \citet{raso19}
suggest that only one star, BSS1, is  marginally consistent
with a mass slightly larger than twice the turnoff mass. In this case the fraction of descendants
from triple systems would be 1/23 $\sim 4\%$. 
\citet{Shara1997} measured the mass of a BSS in 47Tuc
and found 1.7$\pm 0.4\,\msun$. They argue that the most likely path for formation
of this star is the (slow) coalescence of a binary formed by two stars of nearly 0.9\,$\msun$
each. 
\citet{Fiorentino2014} have studied the pulsational masses of BSSs in the
Globular Cluster NGC 6541 (metallicity of --1.76, age= 13.25 Gyr). They found BSSs in 
the mass range 1.0--1.1~$\msun$, in agreement with masses deduced from evolutionary
tracks of single stars. \\

To get some further  insight 
on the fraction of triple systems
we decided to look at the
{\it Gaia} Universe Model \citep[herefater GUM,][]{GUM}\footnote{The {\it Gaia} Universe Model
is available through the {\it Gaia} archive 
(\url{https://gea.esac.esa.int/archive/}) and the relevant
documentation can be found at \url{https://gea.esac.esa.int/archive/documentation/GDR3/Data_processing/chap_cu2sim/sec_cu2UM/}}.
The binary frequency and how this is folded into GUM is discussed
 by \citet{arenou11}. 
The fraction of triple systems among solar type stars
is assumed to be 8.4\% based on the results of \citet{1991A&A...248..485D}.
We selected from GUM stars with absolute magnitudes $M_G > 6$, metallicities [Fe/H]$< 0$
and Population $>1$, that amounts to exclude disc stars.
From the color-magnitude we selected the turnoff stars
and found a fraction of binary systems of 31\%,
and 6\% of triple systems. 
The latter value is not
too different from the 4\% that we estimated above for the BSSs descendants
from triple systems in 47\,Tuc. 
The fraction of triple systems is an upper limit to the fraction
of the systems that can be progenitors of a BSSs of large mass.
We must assume the system is hierarchical, otherwise it is unstable. 
We also assume the system
has an inner couple, that at some stage undergoes a common envelope phase.
In order for the outer third star, to be dragged into this common envelope
and give rise to the fusion of the three stars the outer semi-major
axis must be smaller or equal to the radius of the common
envelope.
From the above considerations, both from observations
and from  GUM we conclude that this channel of formation of BSS
cannot exceed a few percent of the BSS population and is unlikely to explain their observed number.\\

In any case it is not only the fraction of triples that needs to be taken
into account, but the fraction of triples that can result in a triple
merger. On this the theoretical study of \citet{toonen22} 
is illuminating, since it shows that the collision of the inner couple
in a destabilised triple system occurs in 13\%--24\% of the cases,
only for 2.4\% of these the tertiary star may also collide
resulting in a fusion of the three stars. We defer the reader on the discussion
on this point in \citet{bonifacio24} who conclude, factoring all possible events,
that at most 0.3\% of the BSS may result from the fusion of three stars.
This is not in contradiction with the suggestion of
\citet{2009ApJ...697.1048P} that triple systems are one of the main channels
for the formation of BSS. In fact they envisage  only the fusion of the inner
couple, with mass less than or equal to twice the TO mass, with the tertiary remaining
as companion of the BSS. This could explain the large fraction of long-period
binaries among BSS.

\end{appendix}

\end{document}